\documentclass{aastex61}
\usepackage{epstopdf}
\usepackage{tikz}
\usepackage{bm}
\usepackage{amsmath}

\newcommand\aastex{AAS\TeX}

\received{xx}
\revised{xx}
\accepted{xx}
\submitjournal{xxx}

\shorttitle{\aastex\ sample article}
\shortauthors{Cai et al.}

\begin{document}

\title{Convectively coupled equatorial trapped waves in stars and planets}

\author[0000-0003-3431-8570]{Tao Cai}
\affil{State Key Laboratory of Lunar and Planetary Sciences, Macau University of Science and Technology, Macau, People's Republic of China \email{tcai@must.edu.mo}}

\author[0000-0003-0454-7890]{Cong Yu}
\affil{School of Physics and Astronomy, Sun Yat-sen University, Zhuhai, 519082, People's Republic of China \email{yucong@mail.sysu.edu.cn}}

\author[0000-0002-8033-2974]{Xing Wei}
\affil{Department of Astronomy, Beijing Normal University, Beijing, People's Republic of China}



\begin{abstract}
In this paper, we have studied the convectively coupled equatorially trapped waves in rotating stars, with and without magnetic field. The equatorial trapped HD and MHD Poincar\'e, Rossby, mixed Rossby-Poincar\'e, and Kelvin waves were identified. The effects of stratification and non-traditional Coriolis force terms have been investigated. When the flow is strongly stratified, the wave frequencies of the convectively coupled model are almost the same as those of shallow water model. However, when the flow is weakly stratified, the wave frequencies are constrained by the buoyancy frequency. The non-traditional Coriolis terms affect the widths and phases of the equatorial waves. The width increases with the increasing non-traditional Coriolis parameter. Phase shift occurs when the non-traditional Coriolis parameter is included. Magnetic effect is significant when the magnetic field is strong. We have applied the model in the solar atmosphere and solar tachocline to explain the Rieger type periodicities. For the solar atmosphere, when magnetic effect is taken into account, we find that the magnetic field should be smaller than $5G$ in the solar photosphere. Otherwise, the Rieger type periodicities can be only attributed to long Rossby waves. For the solar tachocline, we find that magnetic field of the solar tachocline should be smaller than $50kG$ to observe the 160 days Rieger period. In addition, we find that the effect of the non-traditional Coriolis terms is not obvious in the solar photosphere, but its effect on the tachocline is significant.
\end{abstract}




\section{Introduction}
In rotating stars or planets, waves could be trapped in the equatorial region because of the equatorial beta effect. Long period equatorial Rossby waves could possibly be excited, and it probably plays an important role in dissipating energy of large scale fluid motions \citep{vallis93}. It has been long believed that the Rossby waves exist in the Sun. Rossby-like waves have been observed by following the traces of brightpoints in the Sun \citep{mcintosh17}. Recently, global Rossby waves have been unambiguously detected in the shallow subsurface layers of the Sun from the data of the Helioseismic and Magnetic Imager \citep{loptien18}. The global Rossby waves were also confirmed later by \citet{liang19} from the SOHO data and \citet{hanson20} from the GONG++ data. Theoretical analysis on Rossby waves was typically performed based on the shallow water model. \citet{lou00} used a hydrodynamic shallow water model to predict the 150-160 days Rieger periodicity \citep{rieger84} observed in the solar flares. He suggested that the Rieger type periods are closely related to the equatorially trapped Rossby or mixed Rossby-Poincar\'e waves. \citet{zaqarashvili10} used a magnetohydrodynamic shallow water model and suggested that the Rieger type periods could be explained by magnetic Rossby waves in the solar tachocline. The magnetohydrodynamic shallow water model was also used to estimate the dynamo magnetic field strength in the solar interior \citep{gurgenashvili17a} and predict the solar cycle strength of the sunspots \citep{zaqarashvili15}. Low latitude Rossby waves have also been observed in the atmospheres of Jupiter and Saturn. For the Jupiter, \citet{li06} reported that the North Equatorial Belt wave observed in the Cassini Imaging Science Subsystem is connected to the thermal wave observed in the Composite Infrared Spectrometer. \citet{legarreta16} have also observed a large equatorially trapped wave in the equatorial region of Jupiter. For the Saturn, it has been found that the ribbon waves at $42^{\circ}N$ can be explained as barotropic Rossby waves \citep{gunnarson18}. A summary of Rossby waves in astrophysics can be found in \citet{zaqarashvili21}.

The equatorially trapped waves are usually deduced from shallow water models. Most of the shallow water models mentioned above used traditional approximations (only the vertical component of the Coriolis parameter is considered). Analysis including non-traditional Coriolis force terms (terms associated with the horizonal Coriolis parameter) in the beta plane demonstrated that the dynamics could be changed in a fundamental way \citep{gerkema05}. It has been noted in convectively coupled models, the non-traditional Coriolis terms have important effects on the widths and phases of the equatorial trapped waves \citep{fruman09,roundy12}. Apart from these differences, traditional approximations were also found to have significant effect on wave frequency ranges \citep{cai21}. In shallow water approximations, the depth of fluid is assumed to be much smaller the wavelength of perturbation. However, in stars or planets, the depth of fluid can be comparable to the wavelength. In such case, a general convectively coupled model (density stratification is described by buoyancy frequency) is more appropriate.

The relative strength of buoyancy to rotation plays a key role in the phase structure of the equatorially trapped waves. \citet{fruman09} found that the line of constant phase is almost orthogonal to gravity when the buoyant effect dominates, and almost tangent to the rotation vector when the rotational effect dominates. \citet{roundy12} found that substantial meridional tilts and phase shifts appear when the Coriolis force is strong. \citet{fruman09} and \citet{roundy12} focused their discussions with the convectively couple model on the equatorially trapped waves in the Earth atmosphere. In stars or planets, the depths of atmospheres and rotation rates can be much different. In this paper, we will use the convectively coupled model to discuss the equatorially trapped waves in the stars and planets.

In the Earth atmosphere, the magnetic field is weak and its effect on equatorially trapped waves is negligible. Thus, it is sufficient to describe these waves by a hydrodynamic (HD) model. However, in a star or planet, the magnetic field can be very strong and its effect can be important. When magnetic field is presented, a magneto hydrodynamic (MHD) model instead of HD model should be used. In this paper, we have developed a convectively coupled MHD model for equatorial waves. From this model, we have deduced MHD Poincar\'e waves, Rossby waves, mixed Poincar\'e-Rossby waves, and Kelvin waves. When the magnetic field vanishes, these MHD waves are degenerated into HD waves. This MHD model has been applied to study waves in the solar atmosphere and tachocline. The differences between the MHD and HD models are also discussed.

\section{The method}
A useful approach for the study of equatorial waves is beta approximation, under which the Coriolis parameter is assumed to vary linearly with latitude. To reduce the mathematical complexity, we also assume that the flow is incompressible so that the Boussinesq approximation can be used. The Boussinesq approximation assumes that the density does not vary much throughout the computational domain. It is approximately valid in our case since we only consider a thin layer in a star or planet. In the beta plane, the linearised magnetic hydrodynamic equations for rotating Boussinesq flow are
\begin{eqnarray}
&&\bm{\nabla} \bm{\cdot} \bm{u} = 0~,\label{eq1}\\
&&\partial_{t}\bm{u}  + \bm{f} \bm{\times} \bm{u}  +\bm{\nabla} p-b \hat{\bm{z}}+\frac{1}{4\pi\rho_{0}} \bm{B}_{0}\bm{\times}(\bm{\nabla \times \bm{B_{1}}}) = 0~,\label{eq2}\\
&&\partial_{t} b+ N^2 \bm{u}\cdot \hat{\bm{z}} = 0~,\label{eq3}\\
&& \frac{\partial \bm{B_{1}}}{\partial t}=\bm{\nabla} \bm{\times} (\bm{u}\bm{\times}\bm{B_{0}})~,\label{eq4}\\
&& \nabla \cdot \bm{B_{1}}=0~.\label{eq5}
\end{eqnarray}
where $\bm{u}=(u,v,w)$ is the velocity; $\bm{f}=(0,\tilde{f},f+\beta y)$, with the traditional Coriolis parameter $f=2\Omega\cos\theta$ and the non-traditional Coriolis parameter $\tilde{f}=2\Omega\sin\theta$; $\beta=2\Omega R^{-1}\sin\theta$ is the linear variation rate of the Coriolis parameter along the meridional direction, $R$ is the stellar or planetary radius, $\Omega$ is the rotation rate, and $\theta$ is the colatitude of the beta plane, $p$ is the modified pressure perturbation (pressure scaled by the constant background density); $b$ is the buoyancy; $\bm{B_{0}}$ is the uniform background magnetic field; $\bm{B_{1}}$ is the perturbation of magnetic field; $\rho_{0}$ is the uniform background density; $N^2$ is the square of buoyancy frequency; $\hat{\bm{x}}$, $\hat{\bm{y}}$, and $\hat{\bm{z}}$ are the unit vectors in the zonal, meridional, and vertical directions, respectively; and the operator $\bm{\nabla}$ denotes $\bm {\nabla}=(\partial_{x},\partial_{y}, \partial_{z})$. In this paper, we assume that the background magnetic field is along the toroidal direction with $\bm{B_{0}}=B_{0}\hat{\bm{x}}$, where $B_{0}$ is a constant. Given this assumption, these equations can be written in the Cartesian coordinates as
\begin{eqnarray}
&&\partial_{x} u+ \partial_{y}v+\partial_{z}w=0~,\label{eq6}\\
&&\partial_{t} u - \beta y v+\tilde{f} w +\partial_{x} p=0~,\label{eq7}\\
&&\partial_{t} v +\beta y u + \partial_{y}p-\frac{1}{4\pi\rho_{0}}B_{0}(\partial_{x}B_{1y}-\partial_{y}B_{1x})=0~,\label{eq8}\\
&&\partial_{t} w -\tilde{f} u + \partial_{z}p-b-\frac{1}{4\pi\rho_{0}}B_{0}(\partial_{x}B_{1z}-\partial_{z}B_{1x})=0~,\label{eq9}\\
&&\partial_{t} b+ N^2 w=0~,\label{eq10}\\
&&\partial_{t} B_{1x}-B_{0}\partial_{x}u=0~,\label{eq11}\\
&&\partial_{t} B_{1y}-B_{0}\partial_{x}v=0~,\label{eq12}\\
&&\partial_{t} B_{1z}-B_{0}\partial_{x}w=0~.\label{eq13}
\end{eqnarray}
We assume that the prognostic variables have wave solutions in the vertical and zonal directions in the forms of
\begin{eqnarray}
(u,v,w,p,b,B_{1x},B_{1y},B_{1z})=(\overline{u},\overline{v},\overline{w},\overline{p},\overline{b},\overline{B}_{1x},\overline{B}_{1y},\overline{B}_{1z})\exp(imx+i\ell z-i\sigma t)~, \label{eq14}
\end{eqnarray}
where $m$ and $\ell$ are the wavenumbers in the zonal and vertical directions, respectively; and $\sigma$ is the wave frequency. Substituting (\ref{eq14}) into (\ref{eq6}-\ref{eq13}), we obtain
\begin{eqnarray}
&&im \overline{u}+ \partial_{y}\overline{v}+i\ell \overline{w}=0~,\label{eq15}\\
&&-i\sigma \overline{u} - \beta y \overline{v}+\tilde{f} \overline{w} +im \overline{p} =0~,\label{eq16}\\
&&-i\sigma \overline{v} +\beta y \overline{u} + \partial_{y}\overline{p}-\frac{1}{4\pi\rho_{0}}B_{0}(im\overline{B}_{1y}-\partial_{y}\overline{B}_{1x})=0~,\label{eq17}\\
&&-i \sigma \overline{w} -\tilde{f} \overline{u} + i\ell \overline{p}-\overline{b}-\frac{1}{4\pi\rho_{0}}B_{0}(im\overline{B}_{1z}-i\ell \overline{B}_{1x})=0~,\label{eq18}\\
&&-i\sigma \overline{b}+ N^2 \overline{w}=0~,\label{eq19}\\
&& -i\sigma \overline{B}_{1x}-imB_{0}u=0,\label{eq20}\\
&& -i\sigma \overline{B}_{1y}-imB_{0}v=0,\label{eq21}\\
&& -i\sigma \overline{B}_{1z}-imB_{0}w=0,\label{eq22}
\end{eqnarray}
From (\ref{eq15}-\ref{eq16}) and (\ref{eq18}-\ref{eq22}), we can solve $u$, $w$ and $p$ with the solutions of
\begin{eqnarray}
&&\overline{w}=i\frac{[\ell(\sigma^2-m^2 v_{a}^2)+im \tilde{f}\sigma]\partial_{y}\overline{v}-\beta y m\ell \sigma \overline{v}}{(\ell^2+m^2)(\sigma^2-m^2v_{a}^2)-m^2N^2}~,\label{eq23}\\
&&\overline{p}=-i\frac{[\sigma(\tilde{f}^2+N^2-\sigma^2+m^2v_{a}^2)+i\tilde{f}\ell m v_{a}^2]\partial_{y}\overline{v} +\beta y [m(\sigma^2-(m^2+\ell^2)v_{a}^2-N^2)+i\ell\sigma\tilde{f}] \overline{v}}{(\ell^2+m^2)(\sigma^2-m^2v_{a}^2)-m^2N^2}~,\label{eq24}\\
&&\overline{u}=i\frac{[m(\sigma^2-m^2v_{a}^2-N^2) -i\ell\sigma\tilde{f}]\partial_{y}\overline{v}+\ell^2\sigma \beta y \overline{v}}{(\ell^2+m^2)(\sigma^2-m^2v_{a}^2)-m^2N^2}~,\label{eq25}
\end{eqnarray}
where $v_{a}=B_{0}/\sqrt{4\pi\rho_{0}}$ is the Alfv\'en speed.

Substituting (\ref{eq23}-\ref{eq25}) into (\ref{eq17}), we obtain a second-order partial differential equation of $v$ with a form of
\begin{eqnarray}
\sigma^2(A_{1}\overline{v}+B_{1}\partial_{y}\overline{v}+C_{1}\partial_{yy}\overline{v})+m^2v_{a}^2(A_{2}\overline{v}+B_{2}\partial_{y}\overline{v}+C_{2}\partial_{yy}\overline{v})=0~,\label{eq26}
\end{eqnarray}
with
\begin{eqnarray}
&& A_{1}=\ell^2\sigma^2+m^2\sigma^2-m^2N^2-\ell^2\beta^2 y^2+\beta m(\sigma^2-N^2)/\sigma+i\beta \ell \tilde{f}~,\label{eq27}\\
&& B_{1}=2i\ell \tilde{f}\beta y~,\label{eq28}\\
&& C_{1}=\tilde{f}^2+N^2-\sigma^2~,\label{eq29}\\
&& A_{2}=-(m^2+\ell^2)(2\sigma^2-m^2v_{a}^2-N^2)-\beta m\sigma-\ell^2N^2~,\label{eq30}\\
&& B_{2}=0~,\label{eq31}\\
&& C_{2}=2\sigma^2-m^2v_{a}^2-N^2~.\label{eq32}
\end{eqnarray}
Letting $A=\sigma^2 A_{1}+m^2 v_{a}^2A_{2}$, $B=\sigma^2 B_{1}+m^2 v_{a}^2B_{2}$, and $C=\sigma^2 C_{1}+m^2 v_{a}^2C_{2}$, we can rewrite the above equation as
\begin{eqnarray}
A\overline{v}+B\partial_{y}\overline{v}+C\partial_{yy}\overline{v}=0~.\label{eq33}
\end{eqnarray}
Assuming $\overline{v}=V(y)\exp(\delta y^2)$, we obtain
\begin{eqnarray}
[A+2\delta yB+(2\delta +4\delta^2 y^2)C]V+(B+4\delta y C)\partial_{y}V+C\partial_{yy}V=0~.\label{eq34}
\end{eqnarray}
Letting $\delta=-B/(4yC)$, we can eliminate the first-order derivative term
\begin{eqnarray}
\partial_{yy}V+(\frac{A}{C}-\frac{B}{2yC}-\frac{B^2}{4C^2})V=0~.\label{eq35}
\end{eqnarray}
In the following section, we deduce HD and MHD equatorial waves based on this model.

\section{The Result}
\subsection{Equatorially trapped HD waves}
For hydrodynamic waves, the terms containing $A_{2}$, $B_{2}$, and $C_{2}$ in (\ref{eq26}) are removed.
Thus, (\ref{eq35}) can be written as \citep{roundy12}
\begin{eqnarray}
\partial_{yy}V-\left(\frac{\ell^2\beta^2 (N^2-\sigma^2) }{(\tilde{f}^2+N^2-\sigma^2)^2}\right)y^2 V+\left(\frac{\ell^2\sigma^2-m^2(N^2-\sigma^2)-\beta m(N^2-\sigma^2)/\sigma}{\tilde{f}^2+N^2-\sigma^2}\right)V=0~.\label{eq36}
\end{eqnarray}
This is a Weber differential equation, and the bounded solution
\begin{eqnarray}
V(y)=V_{0} H_{n}\left[\left(\frac{|\ell|\beta\sqrt{N^2-\sigma^2}}{\tilde{f}^2+N^2-\sigma^2}\right)^{1/2}y\right]\exp\left[-\frac{|\ell|\beta\sqrt{N^2-\sigma^2}}{2(\tilde{f}^2+N^2-\sigma^2)}y^2\right] \label{eq37}
\end{eqnarray}
can be obtained if $N^2-\sigma^2>0$ and
\begin{eqnarray}
\frac{\ell^2\sigma^2-m^2(N^2-\sigma^2)-\beta m(N^2-\sigma^2)/\sigma}{|\ell|\beta \sqrt{N^2-\sigma^2}}=2n+1~,\label{eq38}
\end{eqnarray}
where $V_{0}$ is a constant, $n$ is a non-negative integer and $H_{n}$ is the Hermite polynomial of order $n$. Substituting (\ref{eq37}) back, we obtain the zonal velocity
\begin{eqnarray}
&&v=V_{0}H_{n}\left[\left(\frac{|\ell|\beta\sqrt{N^2-\sigma^2}}{\tilde{f}^2+N^2-\sigma^2}\right)^{1/2}y\right]\exp\left[-\frac{|\ell|\beta(\sqrt{N^2-\sigma^2}+i\tilde{f})}{2(\tilde{f}^2+N^2-\sigma^2)}y^2+imx+ilz-i\sigma t\right]~.\label{eq39}
\end{eqnarray}
From (\ref{eq39}), we see that the non-traditional Coriolis parameter $\tilde{f}$ ($\tilde{f}=2\Omega$ at the equator) is involved in the solutions. It has two major effects on the wave solutions. First, $\tilde{f}$ affects the width of the equatorially trapped wave. It can be justified by the looking at the real part of the exponential rate in (\ref{eq39}). The absolute value of the exponential rate decreases with $\tilde{f}$, thus the width of the equatorial trapped wave is wider if $\tilde{f}$ is included. Second, $\tilde{f}$ affects the phase of the equatorially trapped wave. It can be seen from the imaginary part of the exponential rate in (\ref{eq39}). At fixed $(\ell z-\sigma t)$, the constant phase curves ($mx-\ell\beta\tilde{f}/[2(\tilde{f}^2+N^2-\sigma^2)]y^2=consant$) depend on $\tilde{f}$ and are parabolic on the meridional-zonal plane. Phase shifts also occur among different variables. If $\tilde{f}$ is excluded, we see from (\ref{eq23}-\ref{eq25}) that $\overline{w}$, $\overline{p}$, and $\overline{u}$ are all in quadrature to $\overline{v}$ because the coefficients are pure imaginary. However, their phases are out of quadrature when $\tilde{f}$ is included, since the coefficients are not pure imaginary any more. The shifted phase angles can be calculated from the following solutions
\begin{eqnarray}
&&w=\frac{i V_{0}}{\ell^2\sigma^2+m^2\sigma^2-m^2N^2}\left\{\sigma (\ell\sigma+im\tilde{f})\left(\frac{|\ell|\beta\sqrt{N^2-\sigma^2}}{\tilde{f}^2+N^2-\sigma^2}\right)^{1/2}H'_{n}\left[\left(\frac{|\ell|\beta\sqrt{N^2-\sigma^2}}{\tilde{f}^2+N^2-\sigma^2}\right)^{1/2}y\right]\right.\nonumber\\
&&+\left. \beta y\left(\frac{\ell\sigma (\ell\sigma+im\tilde{f})(\sqrt{N^2-\sigma^2}+i\tilde{f})}{(\tilde{f}^2+N^2-\sigma^2)}-m\ell\sigma \right)H_{n}\left[\left(\frac{|\ell|\beta\sqrt{N^2-\sigma^2}}{\tilde{f}^2+N^2-\sigma^2}\right)^{1/2}y\right]\right\}\nonumber\\
&&\quad \exp\left(-\frac{|\ell|\beta(\sqrt{N^2-\sigma^2}+i\tilde{f})}{2(\tilde{f}^2+N^2-\sigma^2)}y^2+imx+ilz-i\sigma t\right)~,\label{eq40}
\end{eqnarray}
\begin{eqnarray}
&&p=\frac{-i V_{0}}{\ell^2\sigma^2+m^2\sigma^2-m^2N^2}\left\{\sigma(\tilde{f}^2+N^2-\sigma^2)\left(\frac{|\ell|\beta\sqrt{N^2-\sigma^2}}{\tilde{f}^2+N^2-\sigma^2}\right)^{1/2}H'_{n}\left[\left(\frac{|\ell|\beta\sqrt{N^2-\sigma^2}}{\tilde{f}^2+N^2-\sigma^2}\right)^{1/2}y\right]\right.\nonumber\\
&&+\left. \beta y\left(\frac{\ell\sigma(\tilde{f}^2+N^2-\sigma^2)(\sqrt{N^2-\sigma^2}+i\tilde{f})}{(\tilde{f}^2+N^2-\sigma^2)}+m\sigma^2-m N^2+i\ell \sigma \tilde{f}\right)H_{n}\left[\left(\frac{|\ell|\beta\sqrt{N^2-\sigma^2}}{\tilde{f}^2+N^2-\sigma^2}\right)^{1/2}y\right]\right\}\nonumber\\
&&\quad \exp\left(-\frac{|\ell|\beta(\sqrt{N^2-\sigma^2}+i\tilde{f})}{2(\tilde{f}^2+N^2-\sigma^2)}y^2+imx+ilz-i\sigma t\right)~,\label{eq41}
\end{eqnarray}
\begin{eqnarray}
&&u=\frac{i V_{0}}{\ell^2\sigma^2+m^2\sigma^2-m^2N^2}\left\{(m\sigma^2-mN^2-i\ell\sigma \tilde{f})\left(\frac{|\ell|\beta\sqrt{N^2-\sigma^2}}{\tilde{f}^2+N^2-\sigma^2}\right)^{1/2}H'_{n}\left[\left(\frac{|\ell|\beta\sqrt{N^2-\sigma^2}}{\tilde{f}^2+N^2-\sigma^2}\right)^{1/2}y\right]\right.\nonumber\\
&&+\left. \beta y\left(\frac{\ell(m\sigma^2-mN^2-i\ell\sigma \tilde{f})(\sqrt{N^2-\sigma^2}+i\tilde{f})}{(\tilde{f}^2+N^2-\sigma^2)}+\ell^2\sigma\right)H_{n}\left[\left(\frac{|\ell|\beta\sqrt{N^2-\sigma^2}}{\tilde{f}^2+N^2-\sigma^2}\right)^{1/2}y\right]\right\}\nonumber\\
&&\quad \exp\left(-\frac{|\ell|\beta(\sqrt{N^2-\sigma^2}+i\tilde{f})}{2(\tilde{f}^2+N^2-\sigma^2)}y^2+imx+ilz-i\sigma t\right)~.\label{eq42}
\end{eqnarray}

Now we discuss the dispersion relations of the waves. The dispersion relation (\ref{eq38}) can be rewritten as
\begin{eqnarray}
\sigma^3-\left[m^2\left(\frac{\sqrt{N^2-\sigma^2}}{|\ell|}\right)^2+(2n+1)\beta\frac{\sqrt{N^2-\sigma^2}}{|\ell|}\right]\sigma-\beta m\left(\frac{\sqrt{N^2-\sigma^2}}{|\ell|}\right)^2=0~\label{eq43}.
\end{eqnarray}
We normalize the frequencies $\sigma$ by $(\beta N/|\ell|)^{1/2}$ and the wave numbers $m$ by $(\beta |\ell|/N)^{1/2}$. After the normalizations, (\ref{eq43}) can be written as
\begin{eqnarray}
\hat{\sigma}^3 -\left[\hat{m}^2 \left(\sqrt{1-\frac{\hat{\sigma}^2}{\hat{N}^2}}\right)^2+(2n+1)\sqrt{1-\frac{\hat{\sigma}^2}{\hat{N}^2}}\right]\hat{\sigma}- \hat{m}\left(\sqrt{1-\frac{\hat{\sigma}^2}{\hat{N}^2}}\right)^2=0~,\label{eq44}
\end{eqnarray}
where $\hat{\sigma}$ and $\hat{m}$ are the normalized variables, and $\hat{N}=(N |\ell|/\beta)^{1/2}$. This dispersion relation is similar to that obtained in shallow water equations \citep{lou00}, except that the shallow-fluid gravity-wave speed is replaced with $\sqrt{N^2-\sigma^2}/\ell$. When $\hat{\sigma} \ll \hat{N}$, it degenerates into the dispersion relation of the shallow water model
\begin{eqnarray}
\hat{\sigma}^3 -\left[\hat{m}^2+(2n+1)\right]\hat{\sigma}- \hat{m}=0~.\label{eq45}
\end{eqnarray}

Analogy to shallow water equations, different kinds of waves can be obtained if appropriate approximations are made. In the high frequency region, the last term in (\ref{eq44}) can be discarded and we obtain the solution
\begin{eqnarray}
\hat{\sigma}=\pm\left[\hat{m}^2\left(1-\frac{\hat{\sigma}^2}{\hat{N}^2}\right)+(2n+1)\sqrt{1-\frac{\hat{\sigma}^2}{\hat{N}^2}}\right]^{1/2} \label{eq46}
\end{eqnarray}
which corresponds to the frequencies of {\it the equatorially trapped Poincar\'e waves}. In the low frequency region, the first term in (\ref{eq44}) can be discarded and we obtain the solution
\begin{eqnarray}
\hat{\sigma}=-\frac{\hat{m}\sqrt{1-\frac{\hat{\sigma}^2}{\hat{N}^2}}}{\hat{m}^2\sqrt{1-\frac{\hat{\sigma}^2}{\hat{N}^2}}+(2n+1)} \label{eq47}
\end{eqnarray}
which corresponds to the frequency of {\it the Rossby wave}. The exact solutions of (\ref{eq44}) can be found for the special case $n=0$. In such case, the solutions are
\begin{eqnarray}
\hat{\sigma}=\frac{\hat{m}}{2}\sqrt{1-\frac{\hat{\sigma}^2}{\hat{N}^2}} \pm \left[ \left(\frac{\hat{m}}{2}\sqrt{1-\frac{\hat{\sigma}^2}{\hat{N}^2}}\right)^2 +\sqrt{1-\frac{\tilde{\sigma}^2}{\tilde{N}^2}} \right]^{1/2} \label{eq48}
\end{eqnarray}
which corresponds to the frequencies of {\it the mixed Rossby-Poincar\'e waves}. Note that the root $\sigma=-{m\sqrt{N^2-\sigma^2}}/{|\ell|}$ has to be rejected for a reasonable solution of the zonal velocity \citep{matsuno66}. Another special case can be deduced if we set the meridional velocity $v=0$ and $n=-1$. Substituting $n=-1$ into (\ref{eq44}), we obtain
\begin{eqnarray}
\hat{\sigma}=\left[ \frac{\hat{N^2}}{1+(\hat{N}/\hat{m})^2} \right]^{1/2} \label{eq49}
\end{eqnarray}
which corresponds to {\it the Kelvin waves}. Note that the negative root has to be rejected for a bounded solution of zonal velocity \citep{matsuno66}.

In these relations, the normalized variables are
\begin{eqnarray}
&&\hat{N}=\left(\frac{N |\ell|}{\beta}\right)^{1/2}=\left( \frac{N}{2\Omega}\right)^{1/2}\left(\frac{2\pi R}{\lambda_{\ell}}\right)^{1/2}=\left( \frac{N}{2\Omega}\right)^{1/2}\left(\frac{R}{H}\right)^{1/2}\left(\frac{2\pi H}{\lambda_{\ell}}\right)^{1/2}~,\\
&&\hat{m}=\left(\frac{m^2 N}{\beta |\ell|}\right)^{1/2}=\left( \frac{N}{2\Omega}\right)^{1/2}\left( \frac{2\pi R} {\lambda_{\ell}}\right)^{1/2}\left( \frac{\lambda_{\ell}}{\lambda_{m}}\right)=\hat{N}\left( \frac{\lambda_{\ell}}{\lambda_{m}}\right)~,
\end{eqnarray}
where $H$ is the height of the wave depth, and $\lambda_{\ell}$ and $\lambda_{m}$ are the wavelengths in the vertical and zonal directions, respectively. $N/(2\Omega)$ measures the degree of the stratification. A large (small) value of $N/(2\Omega)$ indicates the stable layer is strongly (weakly) stratified. $(R/H)$ is the inverse ratio of the wave depth to the spherical radius. $(2\pi H/\lambda_{\ell})$ measures the wave deepness, which is the ratio of the wave depth to the wavelength. In shallow water, this value is usually assumed to be small. $(\lambda_{\ell}/\lambda_{m})$ is the vertical-horizonal wavelength ratio. $(2\pi R/\lambda_{\ell})$ is the ratio of the spherical radius to the vertical wavelength. The smallest value of $(2\pi R/\lambda_{\ell})$ is $2\pi$, thus a small value of $\hat{N}$ can only be achieved when $N/(2\Omega)$ is small (the stable layer is weakly stratified or the rotation is fast).
\begin{figure}
\gridline{\fig{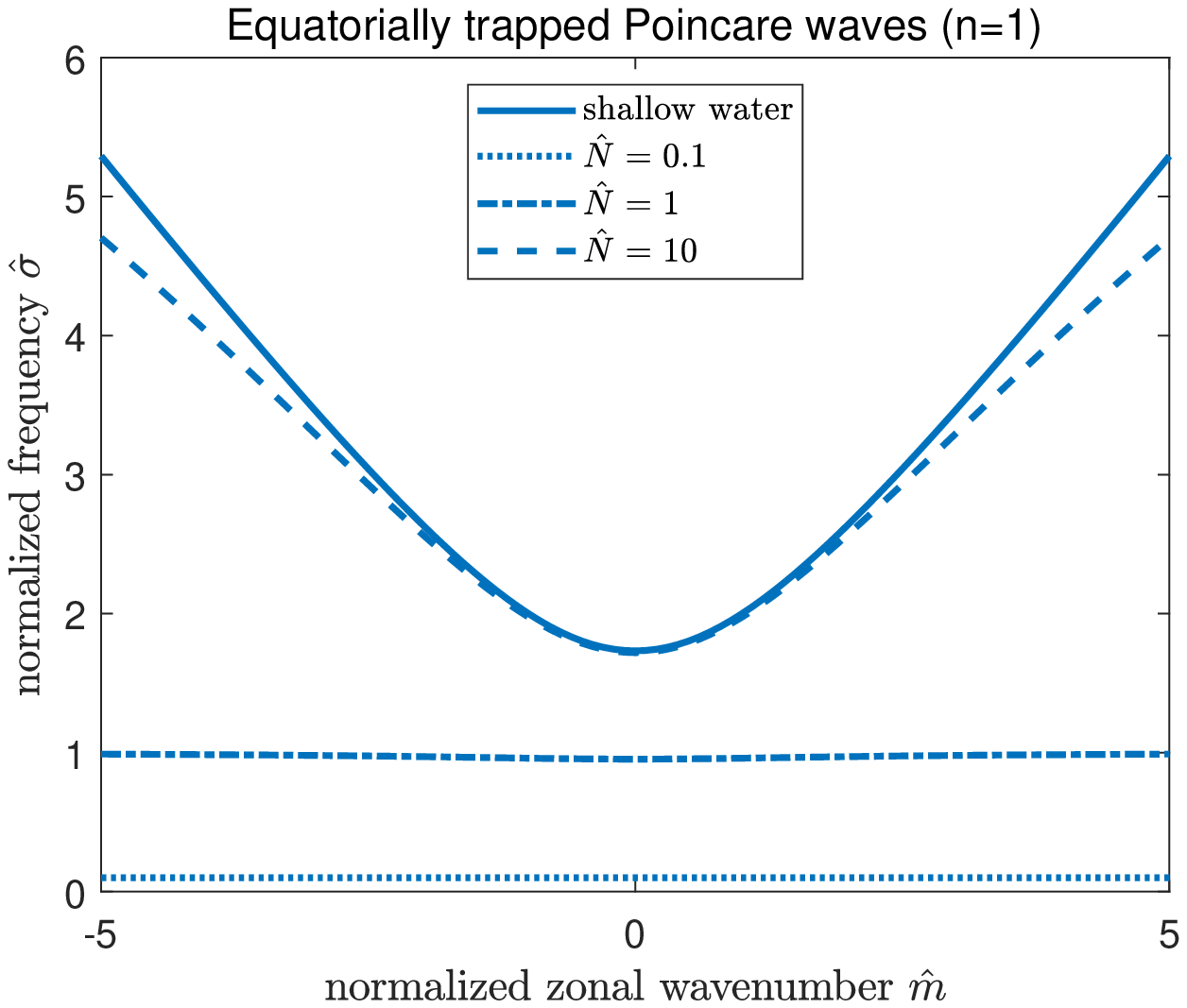}{0.5\textwidth}{(a)}
          \fig{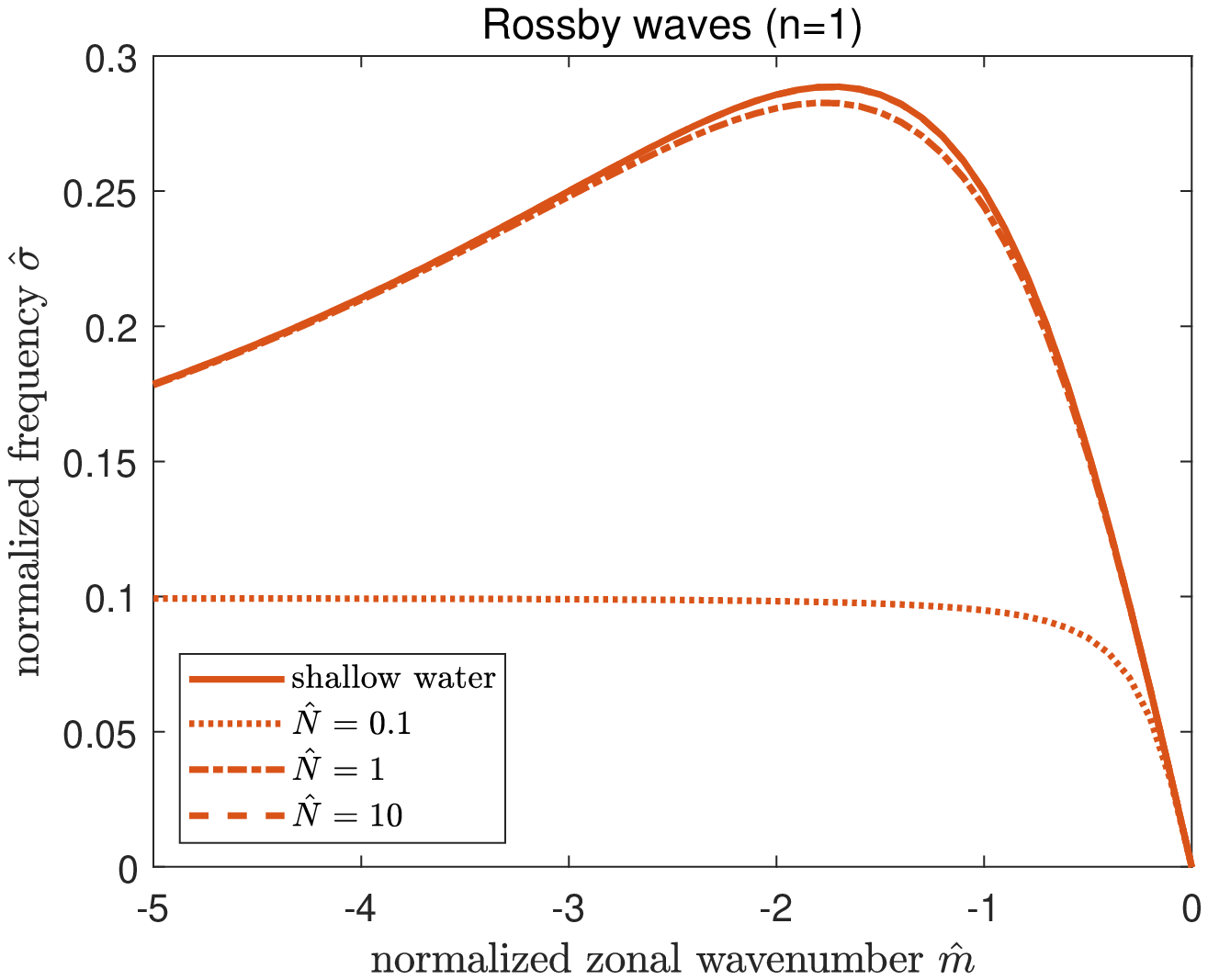}{0.5\textwidth}{(b)}
          }
\gridline{\fig{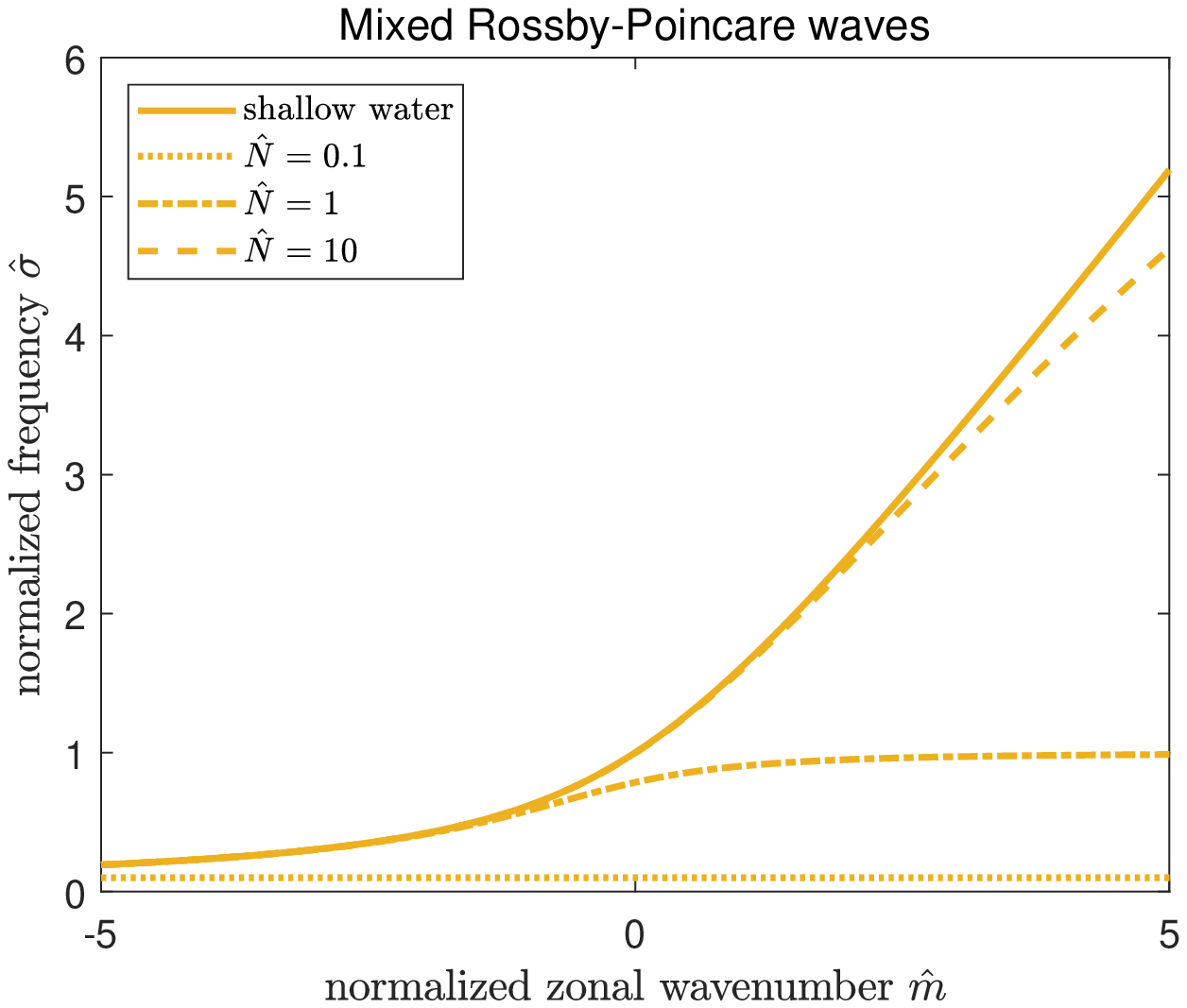}{0.5\textwidth}{(c)}
          \fig{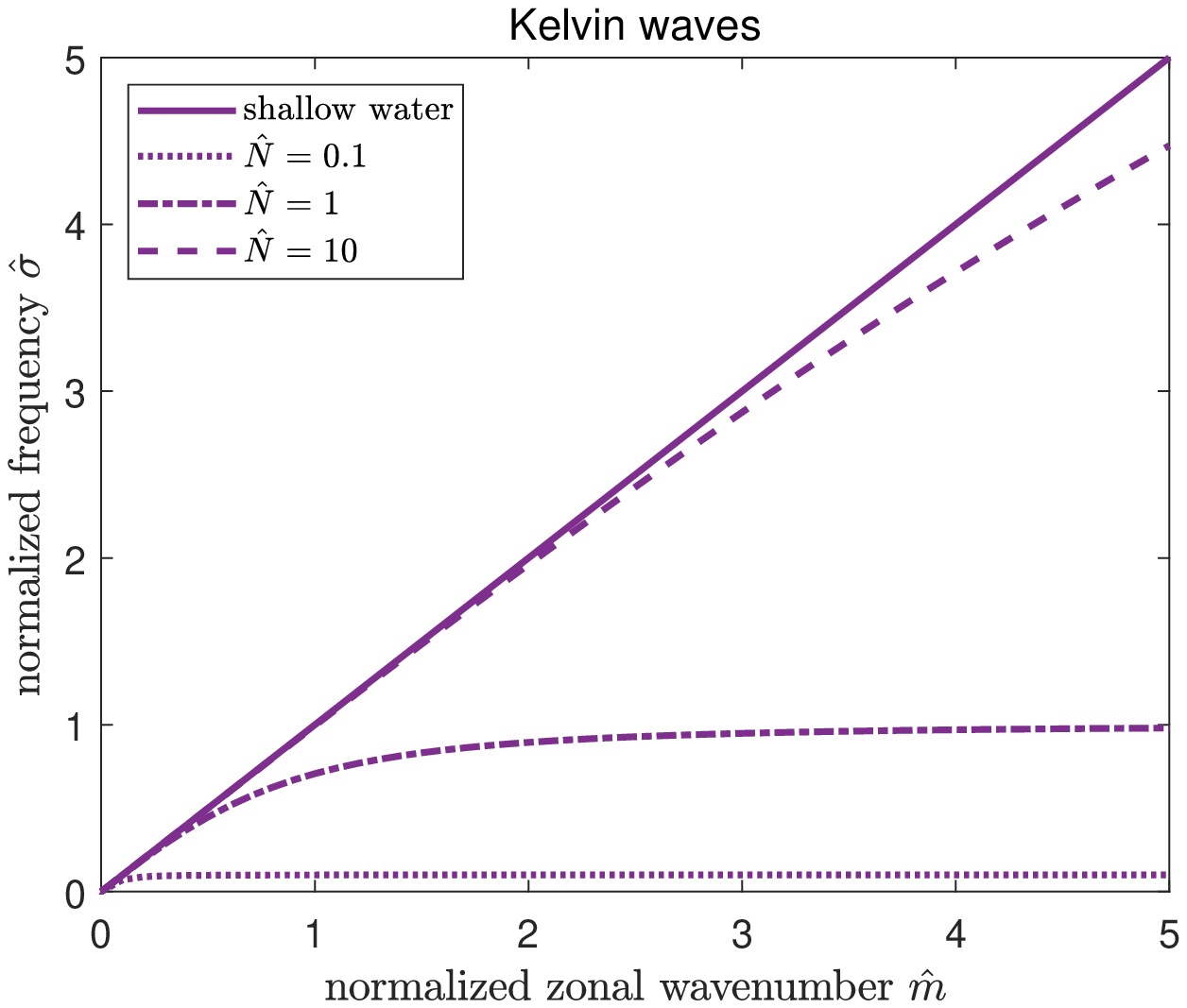}{0.5\textwidth}{(d)}
          }
\gridline{\fig{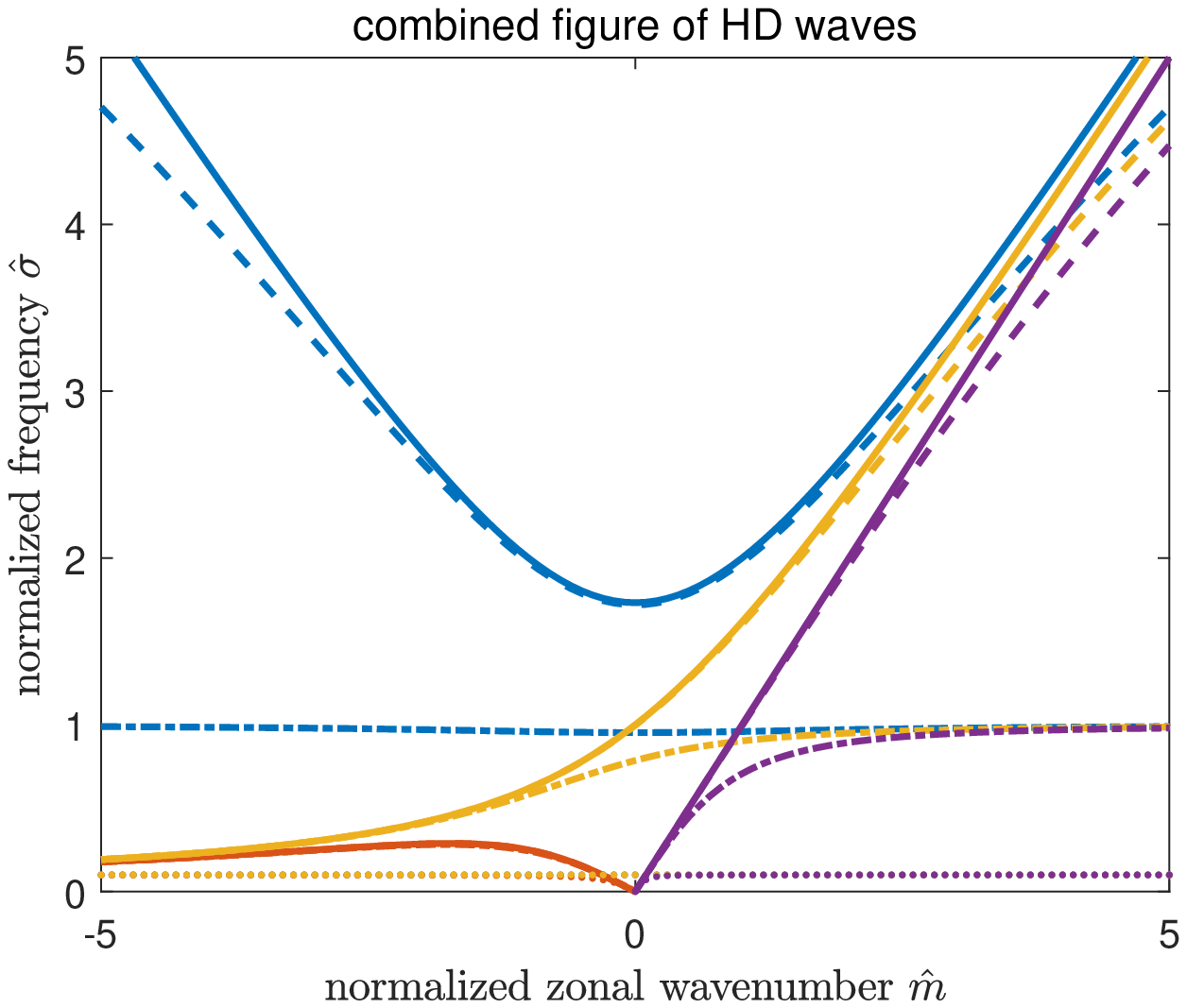}{0.5\textwidth}{(e)}
          }
\caption{Normalized frequency $\hat{\sigma}=\sigma/(\beta N/\ell)^{1/2}$ as a function of normalized zonal wavenumber $\hat{m}=m/(\beta\ell/N)^{1/2}$ for different kinds of waves: (a)the equatorially trapped Poincar\'e waves with $n=1$; (b)the Rossby Waves with $n=1$; (c)the mixed Rossby-Poincar\'e waves; (d)the Kelvin Waves. In each panel, the dispersion relations for convectively coupled waves with different normalized values $\hat{N}=0.1,1,10$ are shown. The dispersion of the shallow water wave is shown for reference. The waves are combined and shown in the panel (e).\label{fig:f1}}
\end{figure}

Fig.~\ref{fig:f1} shows the dispersion relations for the convectively coupled equatorially trapped waves with different values of $\hat{N}$. The dispersions of shallow water waves are also shown for reference. From the figure, we see that the degree of stratification has significant effect on the dispersion relation. It is especially true when the stable layer is weakly stratified.
When the stable layer is strongly stratified, the dispersions of the convectively coupled waves (except for Kelvin waves) are close to those of the shallow water waves. The weakly stratified cases are totally different. We find that the wave frequencies $\hat{\sigma}$ of the Poicar\'e, mixed Rossby-Poincar\'e waves, and Kelvin waves approach to $\pm\hat{N}$ when $\hat{N}$ is small. It can be explained by taking approximations on these waves. For the Poincar\'e waves, it can be deduced by ignoring the second term on the r.h.s. of (\ref{eq46}). For the mixed Rossby-Poincar\'e waves, it can be deduced by using the approximation $(1+x)^\alpha\thickapprox 1+\alpha x$ for the second term on the r.h.s. of (\ref{eq48}). For the Kelvin waves, it can be deduced by ignoring the second term in the denominator of the r.h.s. of (\ref{eq49}). For the Rossby waves, the wave frequencies approach to $\pm \hat{N}$ only when $\hat{m}$ is large. The limits $\hat{\sigma} \rightarrow \pm \hat{N}$ are singular points of (\ref{eq38}). To balance the equation, it also requires $\ell\rightarrow 0$ \citep{roundy12}, indicating that the waves are almost two-dimensional. It can also be seen by letting $\sigma^2=N^2$ in (\ref{eq36}), which yields a wave equation of $V(y)$ with meridional wavelengths of $\pm \ell\sigma/\tilde{f}$. Thus the waves could hardly be equatorially trapped if $|\sigma|$ is close to $N$. For the cases with small values of $\hat{N}$, if a wave is equatorially trapped, it should be a Rossby wave or a Kelvin wave.

The above discussion only consider the frequency range $\sigma^2\leq N^2$. Now we discuss whether equatorial trapped waves could survive in the frequency range $\sigma^2>N^2$ or not. In such case, (\ref{eq36}) has the solutions
\begin{eqnarray}
V(y)=V_{0}W\left(-\frac{\ell^2\sigma^2+m^2(\sigma^2-N^2)+\beta m(\sigma^2-N^2)/\sigma}{|\ell|\beta \sqrt{\sigma^2-N^2}},\left(\frac{|\ell| \beta\sqrt{\sigma^2-N^2}}{\tilde{f}^2+N^2-\sigma^2}\right)^{1/2}y\right)
\end{eqnarray}
for $N^2<\sigma^2<\tilde{f}^2+N^2$, and
\begin{eqnarray}
V(y)=V_{0}W\left(\frac{\ell^2\sigma^2+m^2(\sigma^2-N^2)+\beta m(\sigma^2-N^2)/\sigma}{|\ell|\beta \sqrt{\sigma^2-N^2}},\left(\frac{|\ell| \beta\sqrt{\sigma^2-N^2}}{\sigma^2-\tilde{f}^2-N^2}\right)^{1/2}y\right)
\end{eqnarray}
for $\sigma^2>\tilde{f}^2+N^2$, respectively. Here the function $W(a,\xi)$ is one kind of parabolic cylinder function \citep{abramowitz72}. If $a<0$, then $W(a,\xi)$ oscillates and decays away from $\xi=0$. If $a>0$, then $W(a,\xi)$ oscillates and decays away from $\xi=\pm 2\sqrt{a}$. When $|\xi|\gg |a|$, the modulus function of $W(a,\xi)$ decays with a rate of $\xi^{-1/2}$ \citep{abramowitz72}. The decaying rate is slower than that in the case $\sigma^2<N^2$, where the wave decays exponentially. In the large variable limit $|\xi|\gg |a|$, $W(a,\xi)$ has the following asymptotic expansions \citep{abramowitz72}
\begin{eqnarray}
W(a,\xi)=\sqrt{\frac{2(\sqrt{1+e^{\pi a}}-e^{\pi a})}{\xi}}\cos(\frac{1}{4}\xi^2-a\log\xi+\frac{1}{4}\pi+\frac{1}{2}\arg \Gamma(\frac{1}{2}+ia))~\label{eq54}
\end{eqnarray}
for $\xi>0$, and
\begin{eqnarray}
W(a,\xi)=\sqrt{\frac{2(\sqrt{1+e^{\pi a}}+e^{\pi a})}{\xi}}\sin(\frac{1}{4}\xi^2-a\log\xi+\frac{1}{4}\pi+\frac{1}{2}\arg \Gamma(\frac{1}{2}+ia))~\label{eq55}
\end{eqnarray}
for $\xi<0$, respectively. Here $\arg$ and $\Gamma$ are the argument and gamma functions, respectively. From $(\ref{eq54})$ and $(\ref{eq55})$, we observe that the wave solutions are asymmetrical to the equator for the equatorially trapped waves with frequencies higher than $N$. In the convectively unstable zone, $N^2$ is non positive and we always have $\sigma^2>N^2$. Therefore in the convection zone, we expect that the equatorial trapped waves, if existed, would be asymmetrical to the equatorial plane.

\subsection{Equatorially trapped MHD waves}
For equatorial waves in magneto hydrodynamics, (\ref{eq35}) can be written as
\begin{eqnarray}
&&\partial_{yy}V-\left(\frac{\sigma^2\ell^2\beta^2 (\sigma^2-m^2v_{a}^2)(m^2v_{a}^2+N^2-\sigma^2) }{[\sigma^2\tilde{f}^2+(\sigma^2-m^2v_{a}^2)(m^2v_{a}^2+N^2-\sigma^2)]^2}\right)y^2 V\\
&&+\left(\frac{(\sigma^2-m^2v_{a}^2)^2\ell^2-\sigma\beta m(m^2v_{a}^2+N^2-\sigma^2)-m^2(\sigma^2-m^2v_{a}^2)(m^2v_{a}^2+N^2-\sigma^2) }{\sigma^2\tilde{f}^2+(\sigma^2-m^2v_{a}^2)(m^2v_{a}^2+N^2-\sigma^2)}\right)V=0~.
\end{eqnarray}
Similarly, the bounded solution of the above equation can be found under the condition
\begin{eqnarray}
\frac{(\sigma^2-m^2v_{a}^2)^2\ell^2-\sigma\beta m(m^2v_{a}^2+N^2-\sigma^2)-m^2(\sigma^2-m^2v_{a}^2)(m^2v_{a}^2+N^2-\sigma^2)}{|\sigma| |\ell|\beta\sqrt{(\sigma^2-m^2v_{a}^2)(m^2v_{a}^2+N^2-\sigma^2)}}=2n+1~.
\end{eqnarray}
Normalizing $\omega$ by $(\beta N/|\ell|)^{1/2}$, $m$ by $(\beta|\ell|/N)^{1/2}$, and $v_{a}$ by $N/\ell$, the above equation can be written as
\begin{eqnarray}
&&(\hat{\sigma}^2-\hat{m}^2\hat{v}_{a}^2)^2-\hat{m}^2(\hat{\sigma}^2-\hat{m}^2\hat{v}_{a}^2)\left(1-\frac{\hat{\sigma}^2-\hat{m}^2\hat{v}_{a}^2}{\hat{N}^2}\right)-(2n+1)|\hat{\sigma}| (\hat{\sigma}^2-\hat{m}^2\hat{v}_{a}^2)^{1/2}\left(1-\frac{\hat{\sigma}^2-\hat{m}^2\hat{v}_{a}^2}{\hat{N}^2}\right)^{1/2}\nonumber\\
&&-\hat{\sigma}\hat{m}\left(1-\frac{\hat{\sigma}^2-\hat{m}^2\hat{v}_{a}^2}{\hat{N}^2}\right)=0~,
\end{eqnarray}
where $\hat{\sigma}$, $\hat{m}$, $\hat{v_{a}}$ are normalized variables, and $\hat{N}=(N |\ell|/\beta)^{1/2}$.
Following a similar procedure to the HD case, we obtain
\begin{eqnarray}
\hat{\sigma}=\pm\left[\hat{m}^2\left(1-\frac{\hat{\sigma}^2-\hat{m}^2\hat{v}_{a}^2}{\hat{N}^2}\right)+(2n+1)\sqrt{\left(\frac{\hat{\sigma}^2}{\hat{\sigma}^2-\hat{m}^2\hat{v}_{a}^2}\right)\left(1-\frac{\hat{\sigma}^2-\hat{m}^2\hat{v}_{a}^2}{\hat{N}^2}\right)}+\hat{m}^2\hat{v}_{a}^2\right]^{1/2}
\end{eqnarray}
for {\it the equatorially trapped MHD Poincar\'e waves};
\begin{eqnarray}
\hat{\sigma}=-\frac{\hat{m}\left(1-\frac{\hat{\sigma}^2-\hat{m}^2\hat{v}_{a}^2}{\hat{N}^2}\right)^{1/2}}{\hat{m}^2\left(1-\frac{\hat{m}^2\hat{v}_{a}^2}{\hat{\sigma}^2}\right)\left(1-\frac{\hat{\sigma}^2-\hat{m}^2\hat{v}_{a}^2}{\hat{N}^2}\right)^{1/2}+(2n+1)\left(1-\frac{\hat{m}^2\hat{v}_{a}^2}{\hat{\sigma}^2}\right)^{1/2}}
\end{eqnarray}
for {\it the MHD Rossby waves};
\begin{eqnarray}
\hat{\sigma}=\frac{\hat{m}\left(1-\frac{\hat{\sigma}^2-\hat{m}^2\hat{v}_{a}^2}{\hat{N}^2}\right)^{1/2}\pm \sqrt{\hat{m}^2\left(1-\frac{\hat{\sigma}^2-\hat{m}^2\hat{v}_{a}^2}{\hat{N}^2}\right)+4\left(1-\frac{\hat{\sigma}^2-\hat{m}^2\hat{v}_{a}^2}{\hat{N}^2}\right)^{1/2}}}{2(1-\frac{\hat{m}^2\hat{v}_{a}^2}{\hat{\sigma}^2})^{1/2}}
\end{eqnarray}
for {\it the MHD mixed Rossby-Poincar\'e waves}; and
\begin{eqnarray}
\hat{\sigma}=\left[\frac{\hat{N^2}}{1+(\hat{N}/\hat{m})^2}+\hat{m}^2v_{a}^2\right]^{1/2}
\end{eqnarray}
for {\it the MHD Kelvin wave}. At the hydrodynamic limit $\hat{v}_{a}\rightarrow 0$, the MHD waves degenerate to HD waves.

\begin{figure}
\gridline{\fig{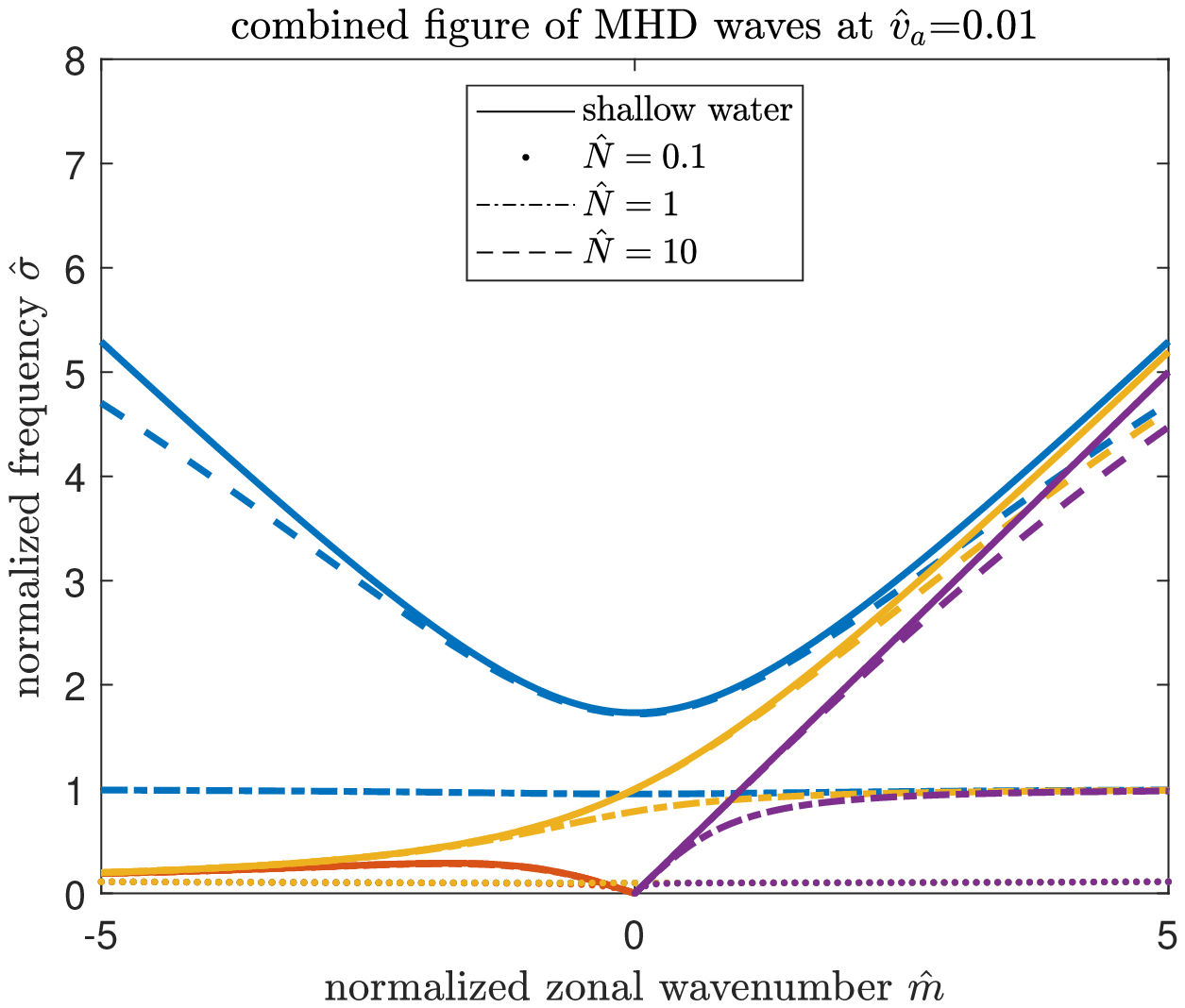}{0.5\textwidth}{(a)}
          \fig{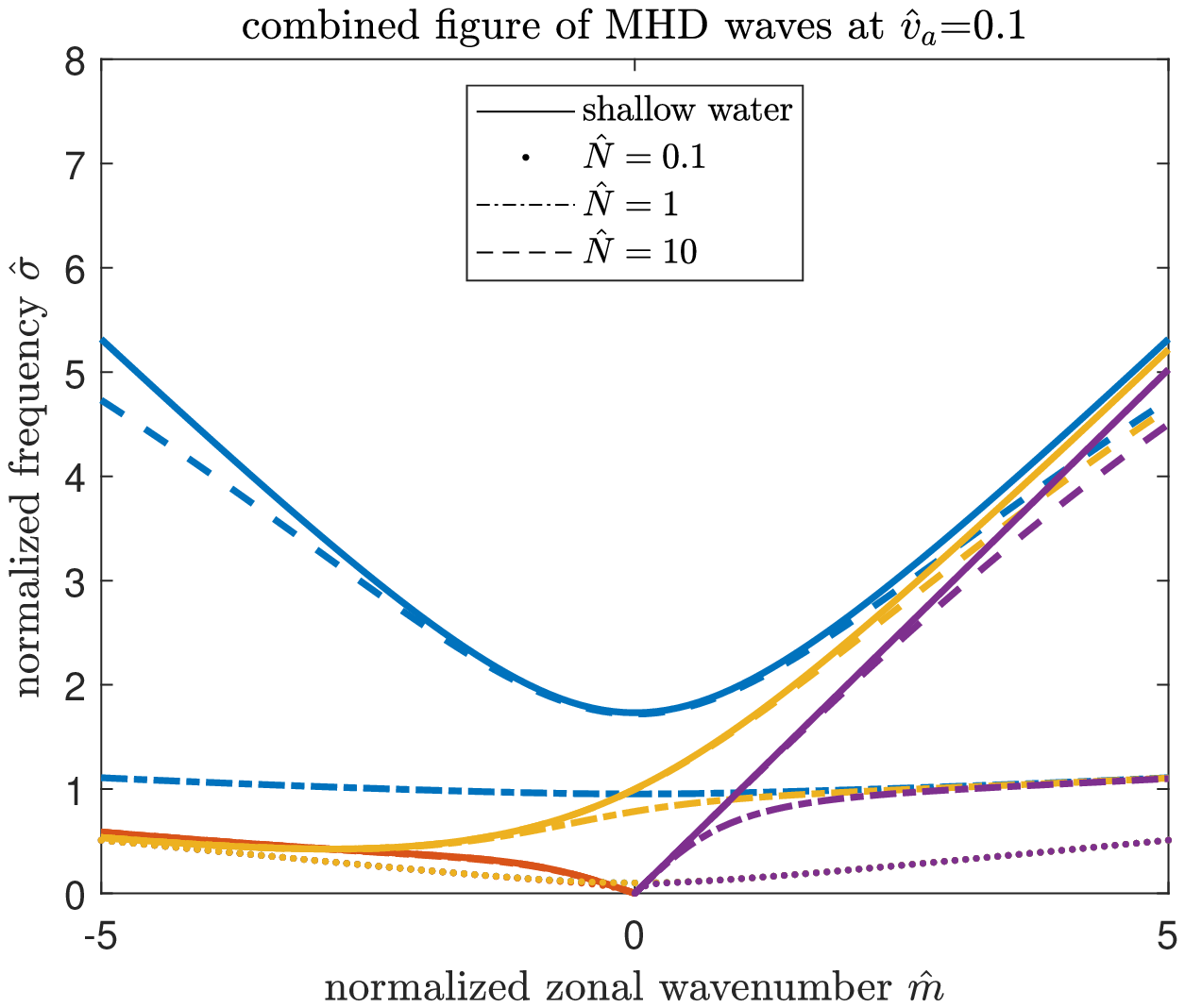}{0.5\textwidth}{(b)}
          }
\gridline{\fig{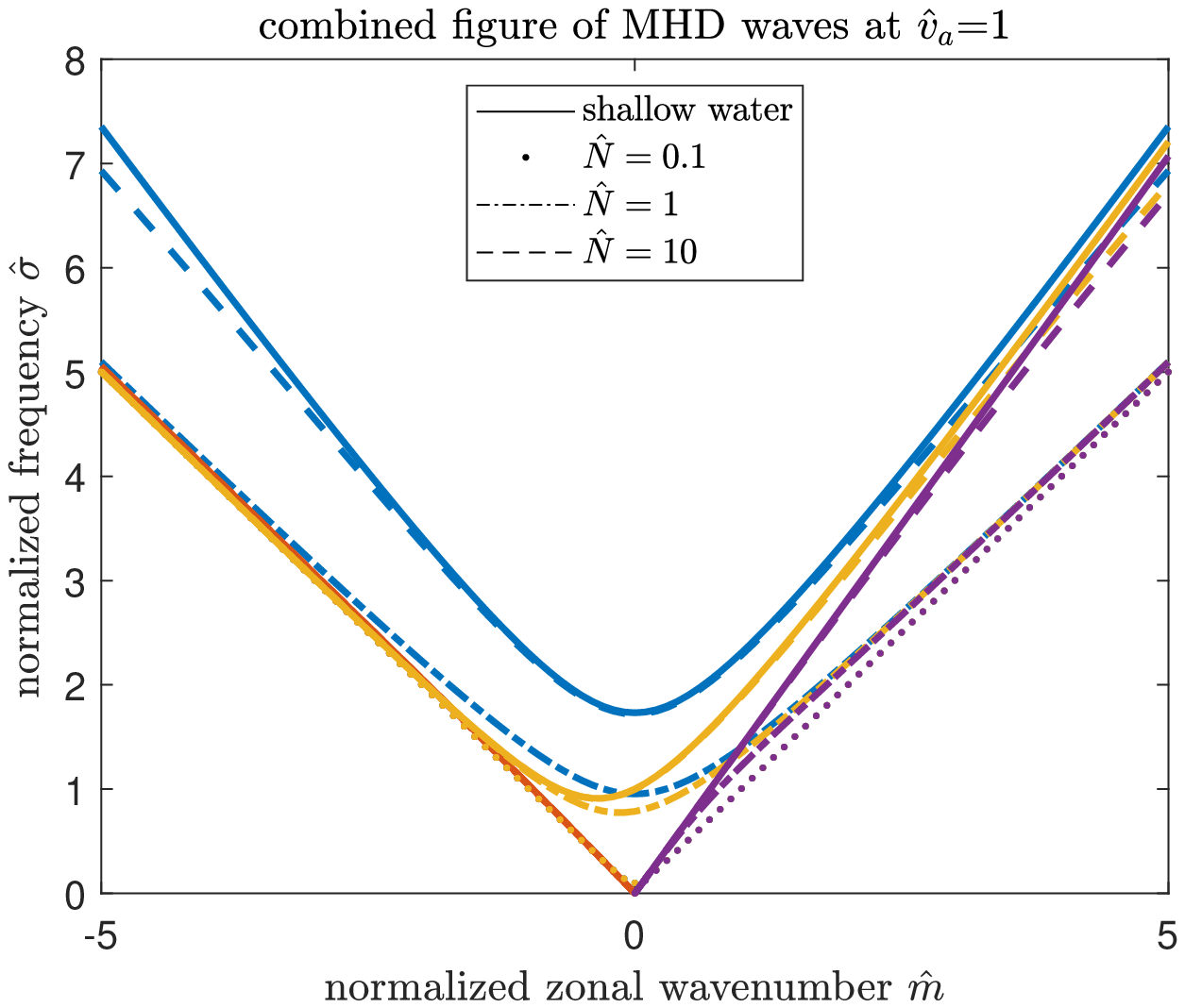}{0.5\textwidth}{(c)}
          \fig{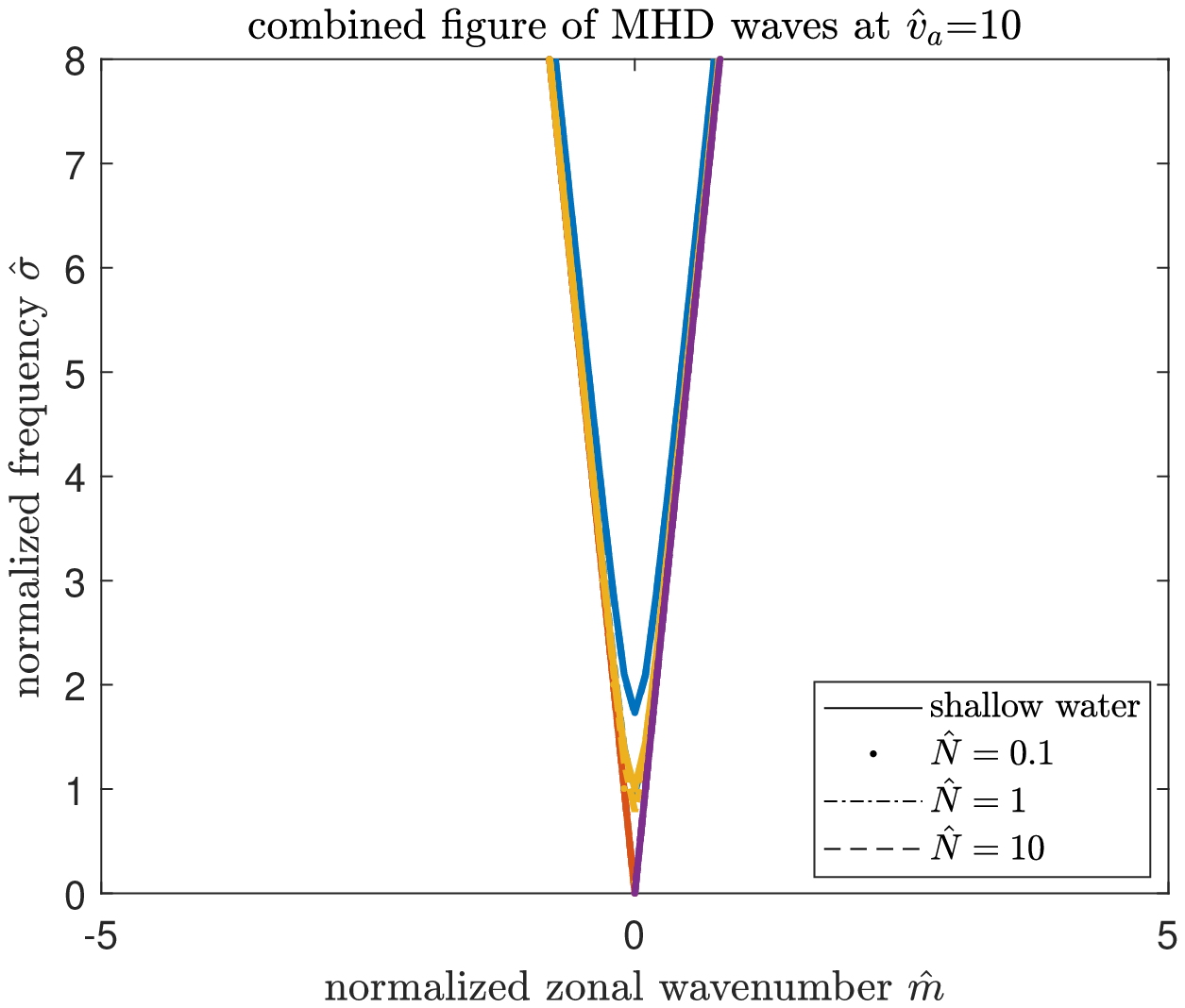}{0.5\textwidth}{(d)}
          }
\caption{MHD waves at different $\hat{v}_{a}$. The blue, brown, yellow, and purple colors represent Poincar\'e waves, Rossby waves, mixed-Poincar\'e waves, and Kelvin waves, respectively. The solid, dotted, dash-dotted, and dashed lines represent shallow water limit, $\hat{N}=0.1$, $\hat{N}=1$, and $\hat{N}=10$, respectively.\label{fig:f2}}
\end{figure}

Fig.~\ref{fig:f2} shows the MHD waves at different $\hat{v}_{a}$ and $\hat{N}$. We notice from Fig.~\ref{fig:f2}(a) that the MHD waves are almost identical to HD waves when the Alfv\'en speed $\hat{v}_{a}$ is small. It indicates that the effect of weak magnetic field on equatorially trapped waves are negligible. Strong magnetic field has significant effect on equatorially trapped waves for both strongly and weakly stratified flows. With a strong magnetic field at $\hat{v}_{a}=10$ (Fig.~\ref{fig:f2}(d)), we see that all the MHD wave speeds approach to the Alfv\'en speed. At $\hat{v}_{a}=0.1$ and $\hat{v}_{a}=1$ (Figs.~\ref{fig:f2}(b) and 2(c)), the different types of waves are modified by the moderate magnetic fields. For both weakly and strongly stratified flow, the equatorially trapped HD wave frequencies are bounded by Brunt-V\"ais\"al\"a frequency $\hat{N}$. The equatorially trapped MHD waves, on the other hand, are bounded by the modified frequency $\sqrt{\hat{m}^2\hat{v}_{a}^2+\hat{N}^2}$. Thus, the magnetic effect will be significant when $\hat{m}\hat{v_{a}}$ is larger than or comparable to $\hat{N}$.

\section{Application to the Sun}
\subsection{Equatorially trapped waves in solar atmosphere}
Now we discuss the equatorially trapped waves in the solar atmosphere. Rieger periodicity, which is around 150-160 days, has been detected in the solar activities \citep{rieger84} (A summary is given in \citet{carbonell92}). Other Rieger type periodicities at around 128, 102, 78, 51 days have also been detected \citep{bai93}. \citet{lou00} has connected these Rieger type periodicities to the equatorially waves in a shallow water model of solar photosphere. The solar radius is about $R_{\odot}\approx 6.957\times10^{10}cm$, and the rotation rate is about $\Omega_{\odot}\approx 2.8469\times 10^{-6}s^{-1}$. In the equatorial region, $\beta=2\Omega_{\odot} R_{\odot}^{-1}\approx 8.184\times 10^{-17}cm^{-1}s^{-1}$. We assume that the depth of the photosphere is approximately $H=5\times10^{7}cm$. The Brunt-V\"ais\"al\"a frequency $N$ is approximately $0.03s^{-1}$ (taken from the latest solar model of \citep{zhang19}). The Brunt-V\"ais\"al\"a period has a magnitude of minutes. Apparently, it is too short compared with the Rieger periods. Thus it is unnecessary to consider the frequency range $\sigma^2>N^2$, and we will only focus on the frequency range $\sigma^2<N^2$. The vertical wavenumber can be written as $\ell=2\pi\tilde{\ell}/H$, where $\tilde{\ell}$ is an integer. The zonal wavenumber can be written as $m=\tilde{m}/R_{\odot}$, where $\tilde{m}$ is an integer.

\begin{figure}
\gridline{\fig{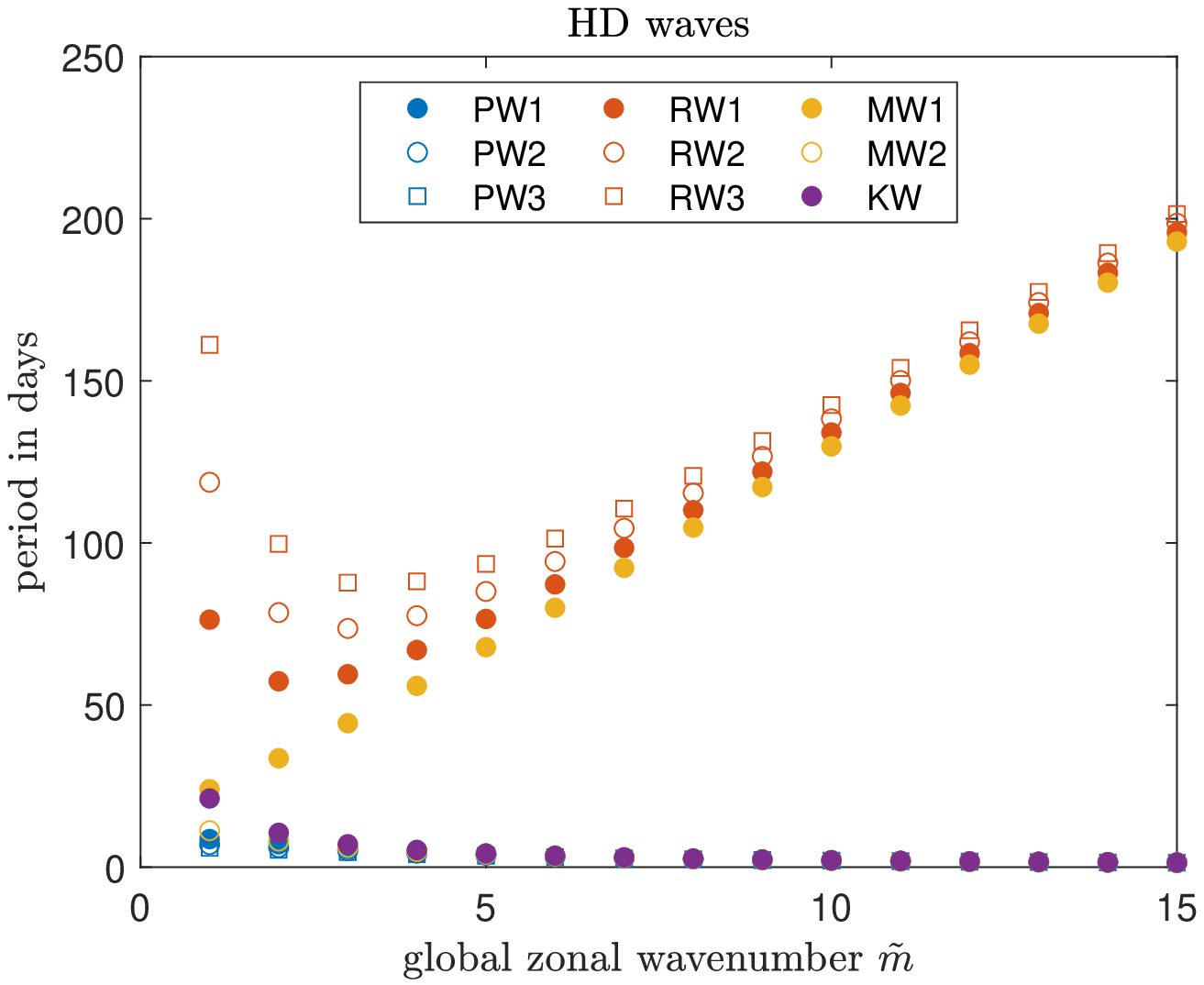}{0.5\textwidth}{(a)}
          \fig{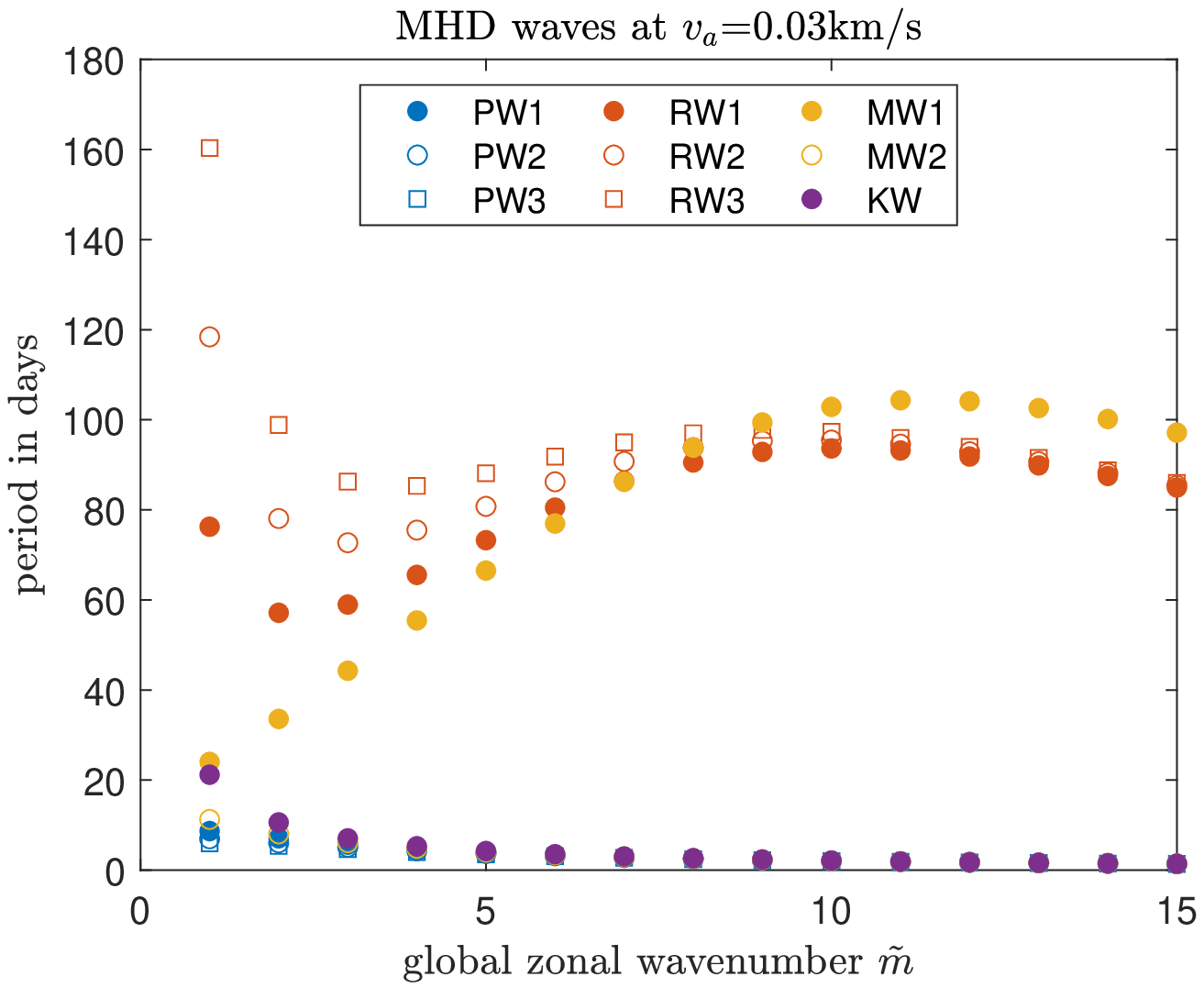}{0.5\textwidth}{(b)}
          }
\gridline{\fig{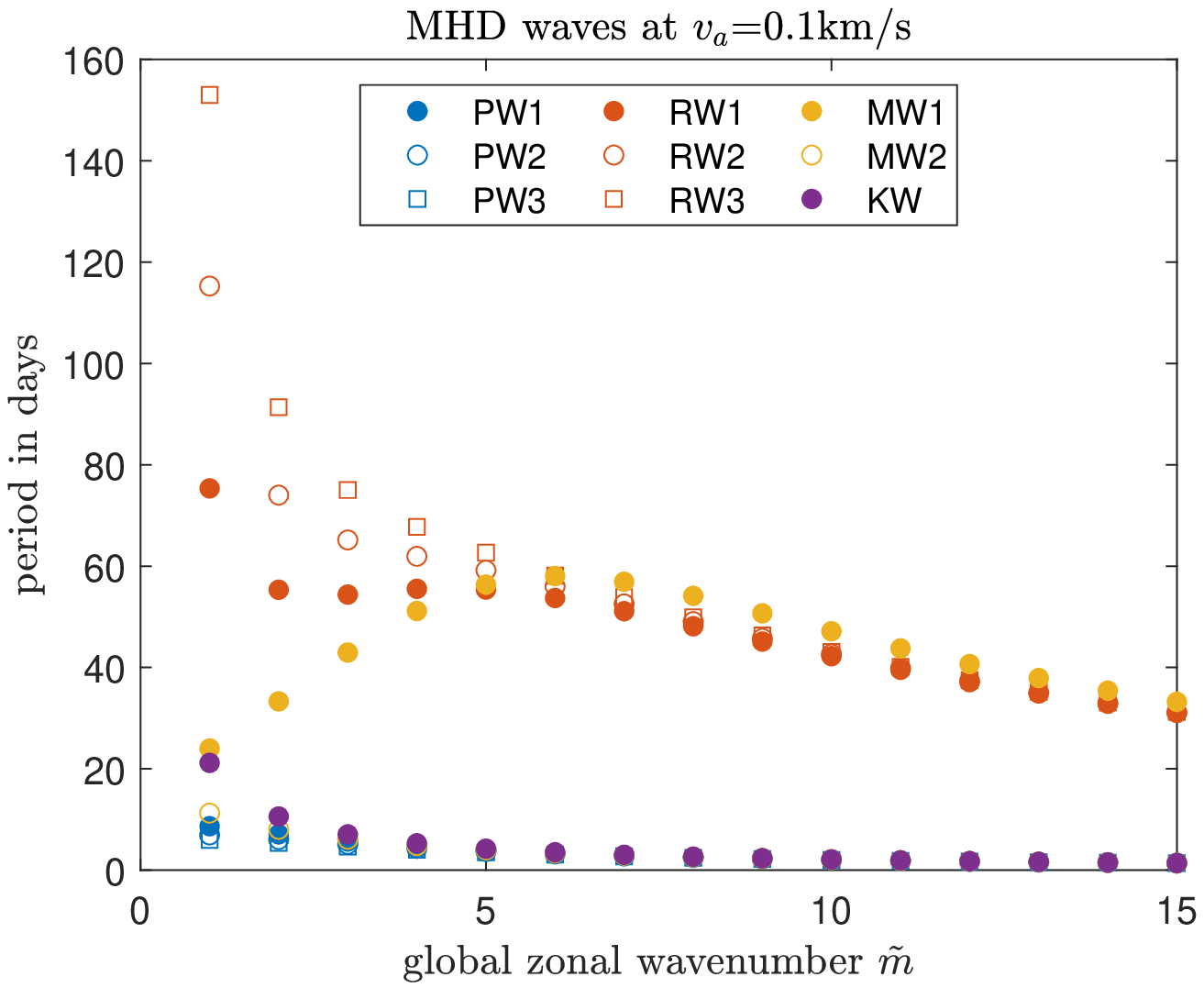}{0.5\textwidth}{(c)}
          \fig{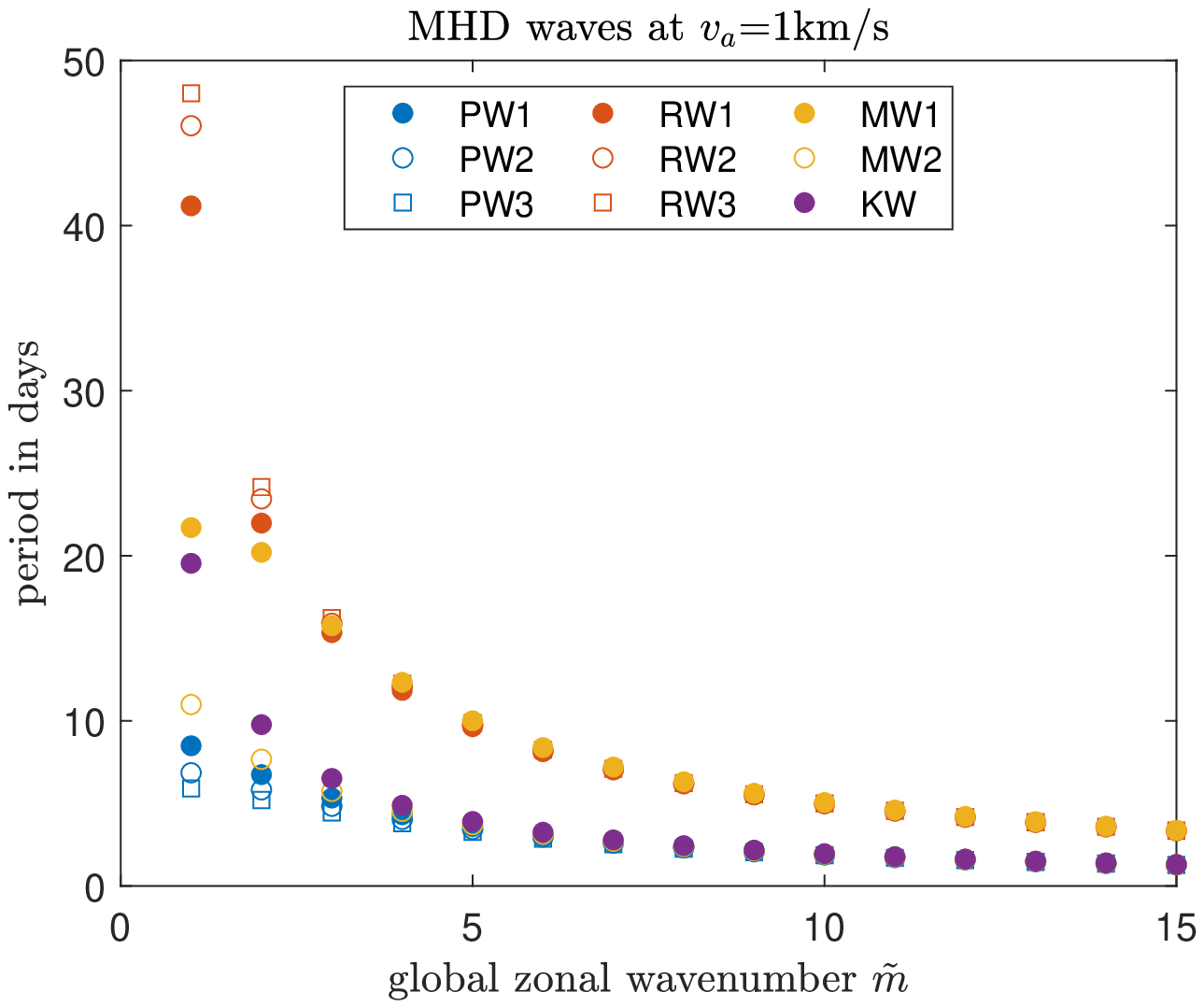}{0.5\textwidth}{(d)}
          }
\caption{Estimated periods in days for different kinds of waves in the solar atmosphere. (a) for HD waves; (b-d) for MHD waves at different Alfv\'en speeds. \label{fig:f3}}
\end{figure}

Fig.~\ref{fig:f3} shows the estimated periods of different types of waves. The panel (a) shows the periods of HD waves, and panels (b-d) show MHD waves at different Alfv\'en speeds. We first consider the HD waves. From Fig.~\ref{fig:f3}(a), we see that the periods of the Rossby and mixed Rossby-Poincar\'e waves have similar magnitudes as the Rieger periodicities. The 150-160 days periodicity corresponds to a global zonal wavenumber $\tilde{m}$ within 11-13 for the Rossby and mixed Rossby-Poincar\'e waves. Other Rieger type periodicities can also be identified. For example,
the 128, 102, 78, 51 days Rieger type periodicities agrees with the periods of the mixed Rossby-Poincar\'e waves at wavenumbers $\tilde{m}$ within $3-10$. In the solar photosphere, compared to the solar rotation rate, the Brunt-V\"ais\"al\"a frequency is large. In such case, the wave periods estimated between the convectively coupled model and the shallow water model have no significant differences. \citet{lou00} used a shallow water model and obtained similar results. At the large $\tilde{m}$ limit, from (\ref{eq47}-\ref{eq48}), we find that the frequencies of Rossby and mixed Rossby-Poincar\'e waves are proportional to $1/{\tilde{m}}$. As a result, the periods predicted in these waves will be proportional to $\tilde{m}$. Fig.~\ref{fig:f3}(a) has verified this trend. Recently, the Rossby wave frequencies have been investigated in \citet{loptien18}, \citet{liang19}, and \citet{hanson20}. Their results revealed that the frequencies are approximately proportional to $1/(\tilde{m}+1)$, which agrees well with the linear analysis in the spherical geometry \citep{saio82}. The difference is caused by the different forms of horizontal Laplacian operators in the Cartesian and spherical geometries \citep{longuet64}. In the Cartesian geometry $\nabla_{h}^2\sim \tilde{m}^2$, while in the spherical geometry $\nabla_{h}^2\sim \tilde{m}(\tilde{m}+1)$. Here we only consider the case with zero node number in the meridional direction. Compared to the spherical geometry, the periods predicted in the Cartesian geometry may lag for a wavenumber.

Wave periods can be significantly modified by magnetic field. The density at the photosphere is about $2\times 10^{-7}g/cm^3$. The values of magnetic field for panels (b-d) are about $5G$, $16G$, and $159G$, respectively. From the figures, we see that the wave periods decrease with increasing toroidal magnetic field. When the strength of magnetic field reaches $159G$, all the wave periods are less than 50 days. Thus, the toroidal magnetic field $B_{0}$ at the solar photosphere cannot be too large. When $B_{0}\approx 16G$, the periods of mixed Rossby-Poincar\'e waves are generally below 100 days. The longer Rieger type period can only be attributed to Rossby wave at small $\tilde{m}$. It is interesting to note that, for high $\tilde{m}$ waves, there is a trend reversal of wave periods when increasing $B_{0}$. For weak $B_{0}$, wave periods with high $\tilde{m}$ generally increase with $\tilde{m}$. However, the trend is reversed for strong $B_{0}$. The trend reversal takes place at a critical value of $B_{0}\approx 5G$ (Fig.~\ref{fig:f2}(b)). As a result, we predict that the toroidal magnetic field should be less than $5G$ at the equatorial region of the Sun, if the intermediate Rieger type periodicities are connected to equatorial waves at high $\tilde{m}$. Otherwise, they must be Rossby waves at low $\tilde{m}$.

\subsection{Equatorially trapped waves in the solar tachocline}
In the above discussion, we consider the equatorially trapped MHD waves in the solar atmosphere. \citet{ballester02} and \citet{ballester04} proposed that there is a casual link between the Rieger type periodicities and the photospheric magnetic flux.
By a shallow water model, \citet{zaqarashvili10} suggested that the Rieger type periodicities are probably driven in the solar tachocline. From previous analyses, we have shown that the buoyancy frequency $N^2$ has important effect on wave frequencies, especially when $N^2$ is small. In the solar tachocline, the buoyancy frequency $N^2$ can be very small in the overshooting layer. Here we discuss the equatorially trapped waves excited in this overshooting layer.

\begin{figure}
\gridline{\fig{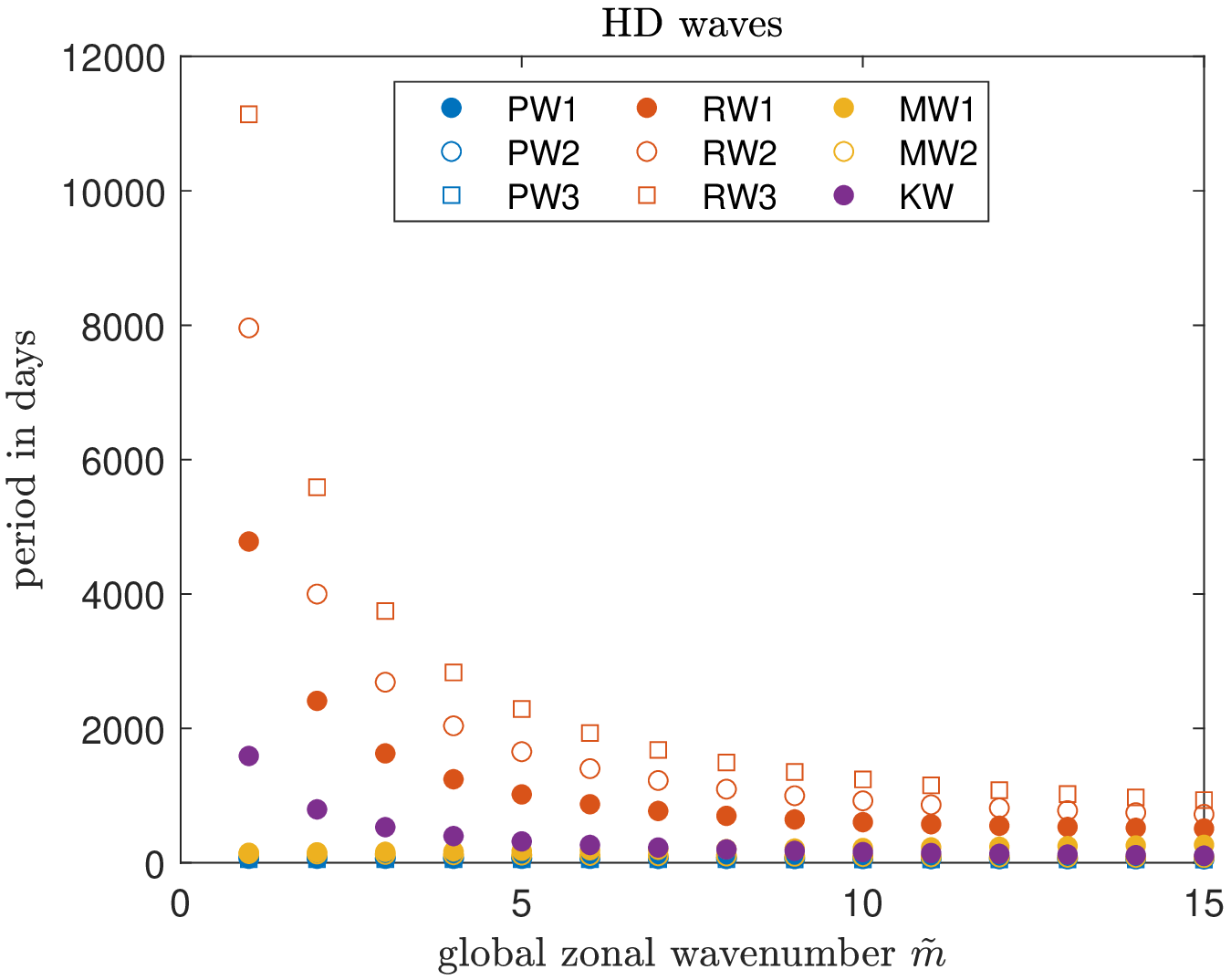}{0.5\textwidth}{(a)}
          \fig{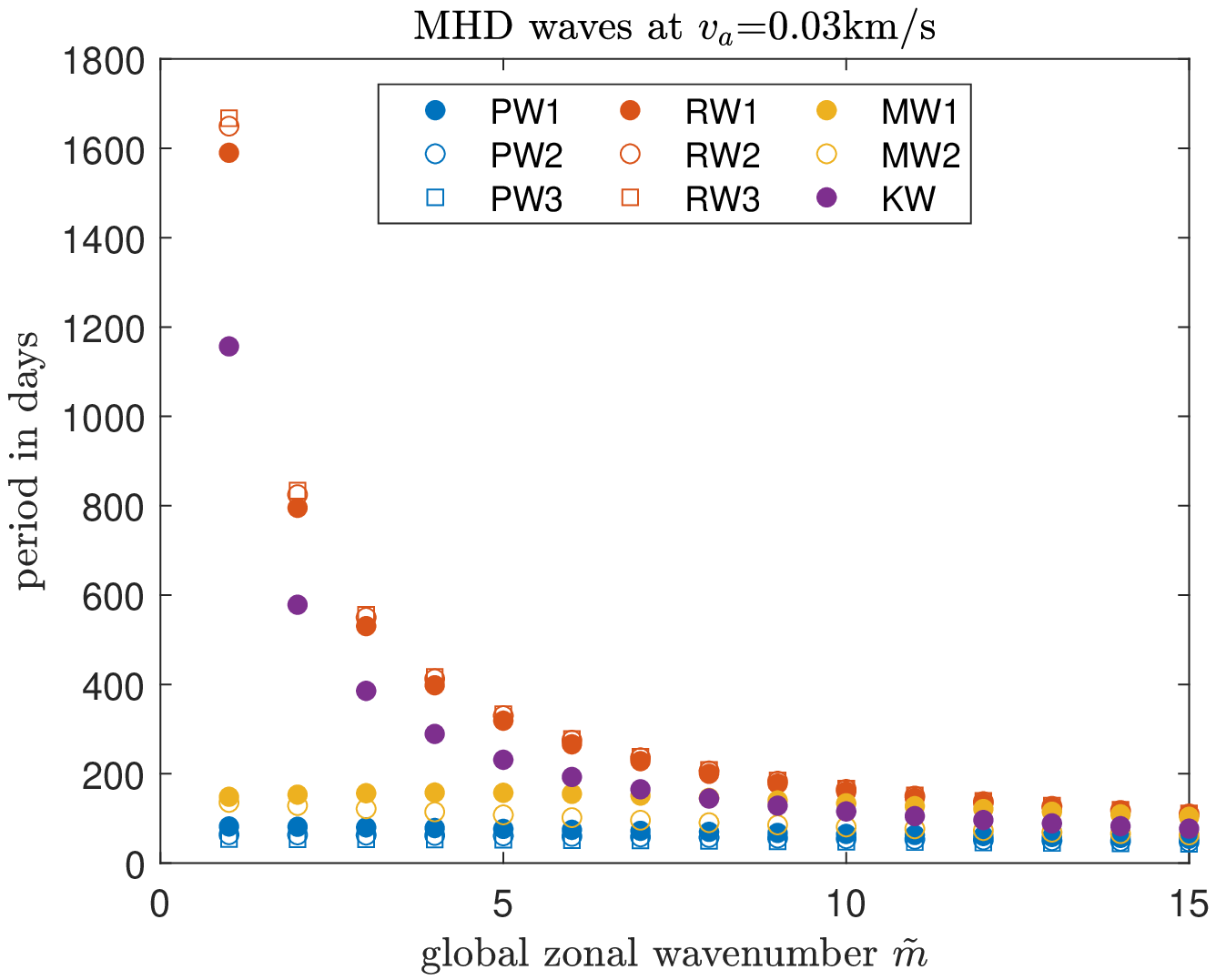}{0.5\textwidth}{(b)}
          }
\gridline{\fig{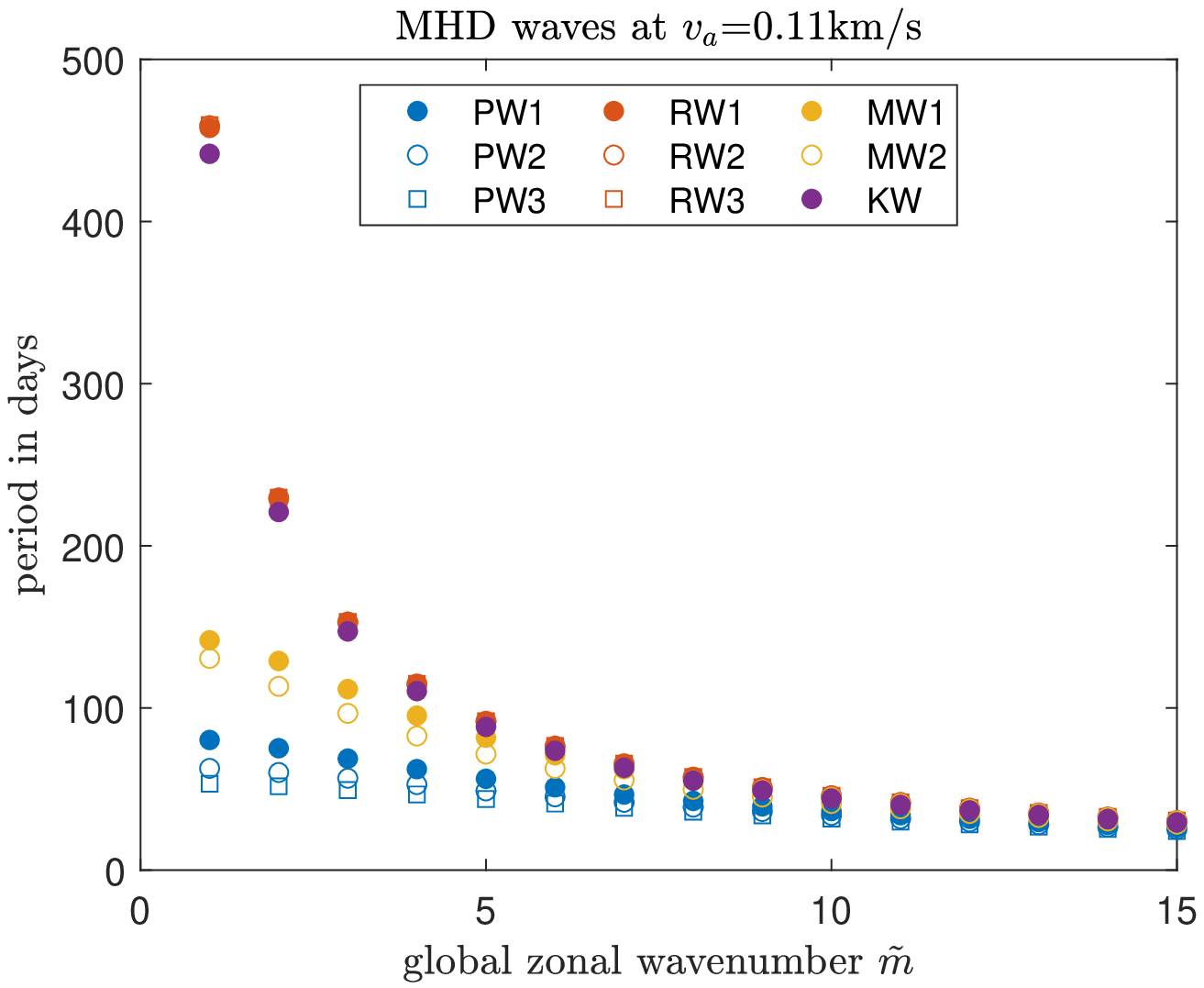}{0.5\textwidth}{(c)}
          \fig{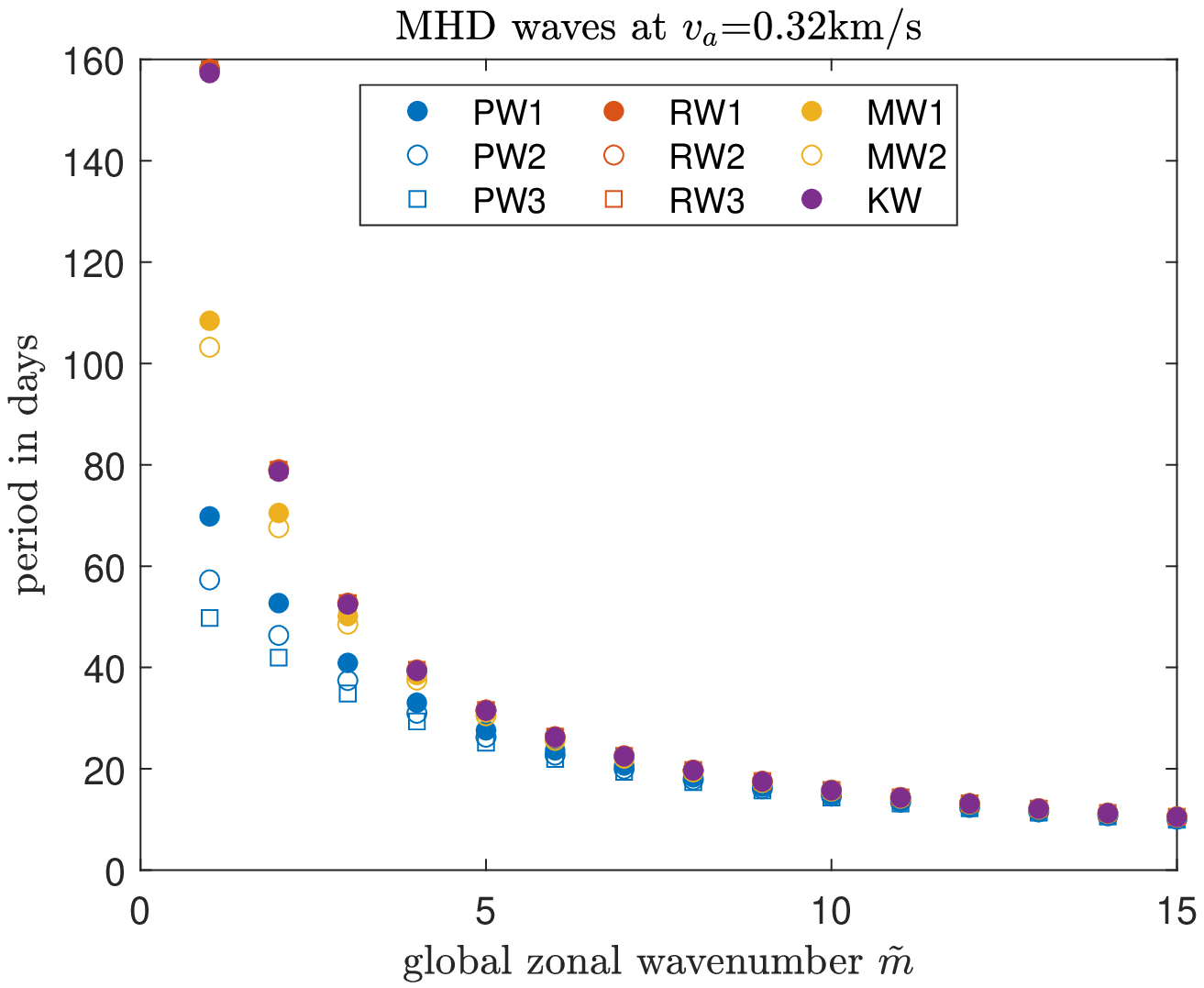}{0.5\textwidth}{(d)}
          }
\caption{Estimated periods in days for different kinds of waves in the solar tachocline. (a) for HD waves; (b-d) for MHD waves at different Alfv\'en speeds.\label{fig:f4}}
\end{figure}

We assume that the overshooting layer has a depth of $2000 km$, and the buoyancy frequency $N^2=1\times 10^{-4} s^{-1}$. The density of this overshooting layer is assumed to be $0.2 g/cm^{-3}$. Fig.~\ref{fig:f4} presents the estimated periods in days for different kinds of waves in the solar tachocline. Fig.~\ref{fig:f4}(a) shows the HD waves, and Figs.\ref{fig:f4}(b-d) shows the MHD waves at different Alfv\'en speeds. For Figs.\ref{fig:f4}(b-d), the strengthes of the magnetic fields are $5kG$, $17kG$, and $50kG$, respectively. From the figures, we see that magnetic field has significant effect on the wave periods. Compared with HD waves, the periods of MHD waves can be significantly reduced. For example, the Rossby wave periods are several thousand days for HD waves, while they are reduced to be about one hundred days for MHD waves when $B_{0}$ is $50kG$. Thus, we conclude that a stronger magnetic field will lead to shorter wave periods. This is consistent with the conclusion drawn by \citet{gachechiladze19}. When $B_{0}$ is greater than $50kG$, all the wave periods are smaller than 160 days. Therefore, we predict that the magnetic field of solar tachocline cannot be stronger than $50kG$.

\subsection{Rotating effect}
\begin{figure}
\gridline{\fig{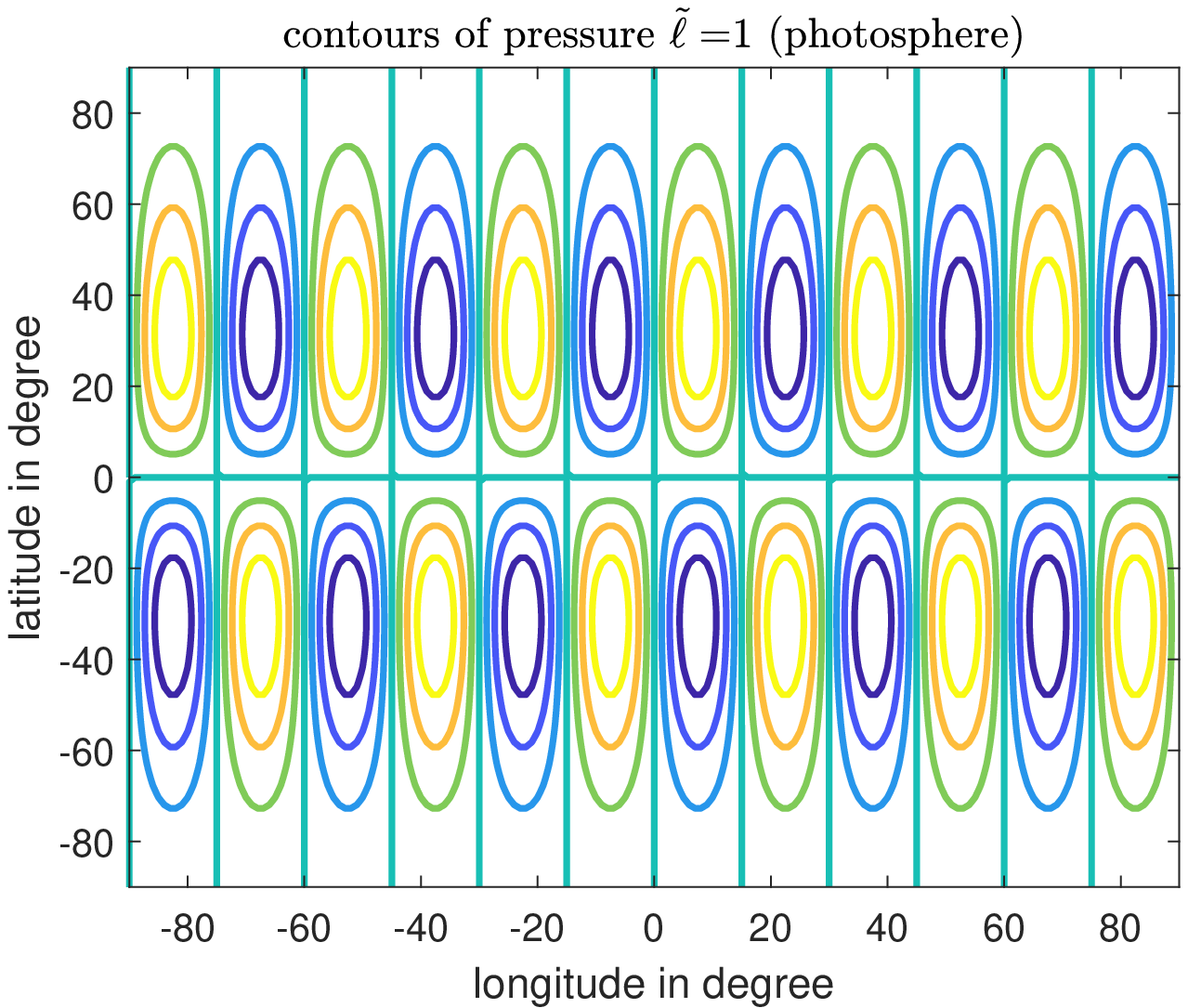}{0.5\textwidth}{(a)}
          \fig{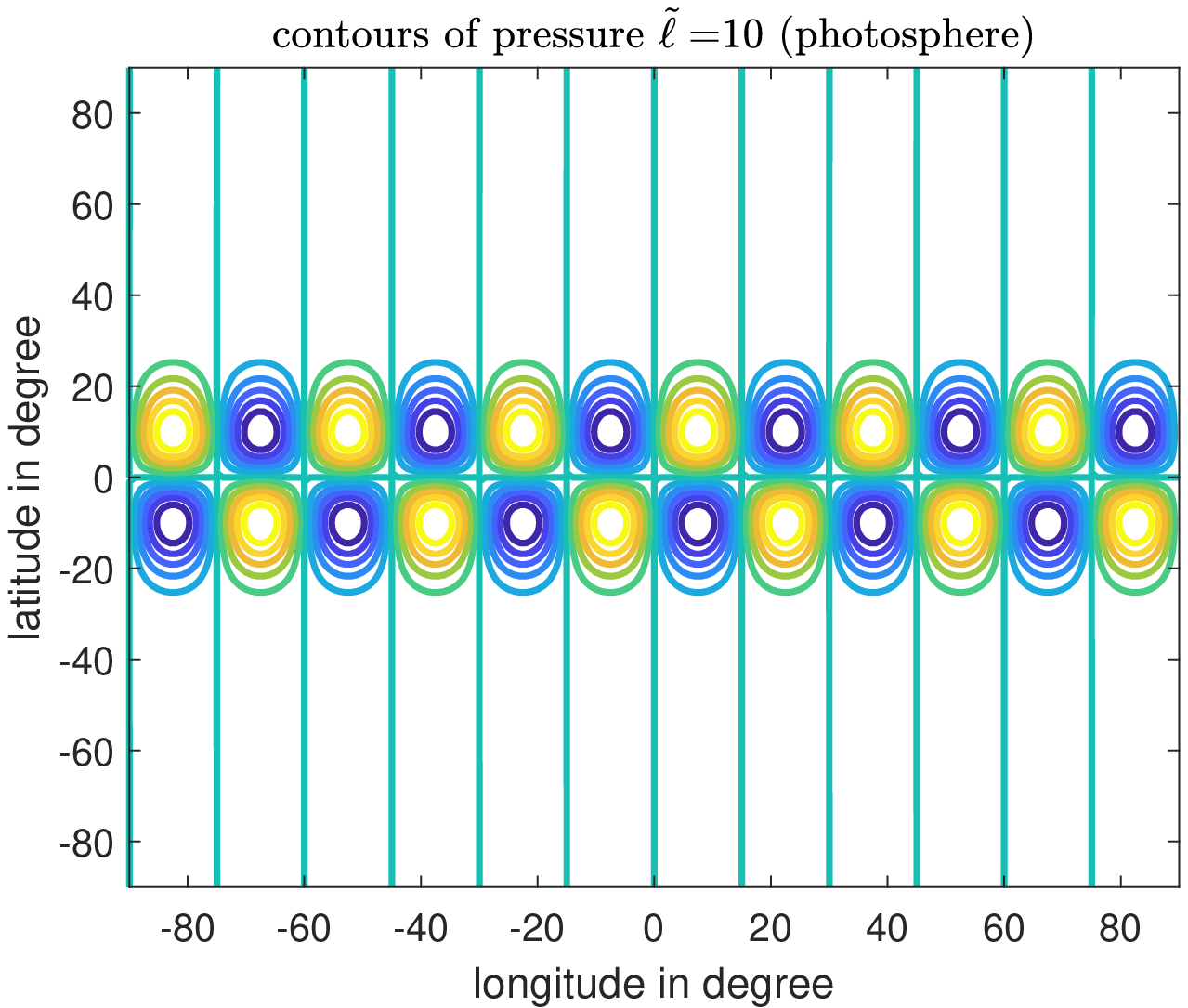}{0.5\textwidth}{(b)}
          }
\gridline{\fig{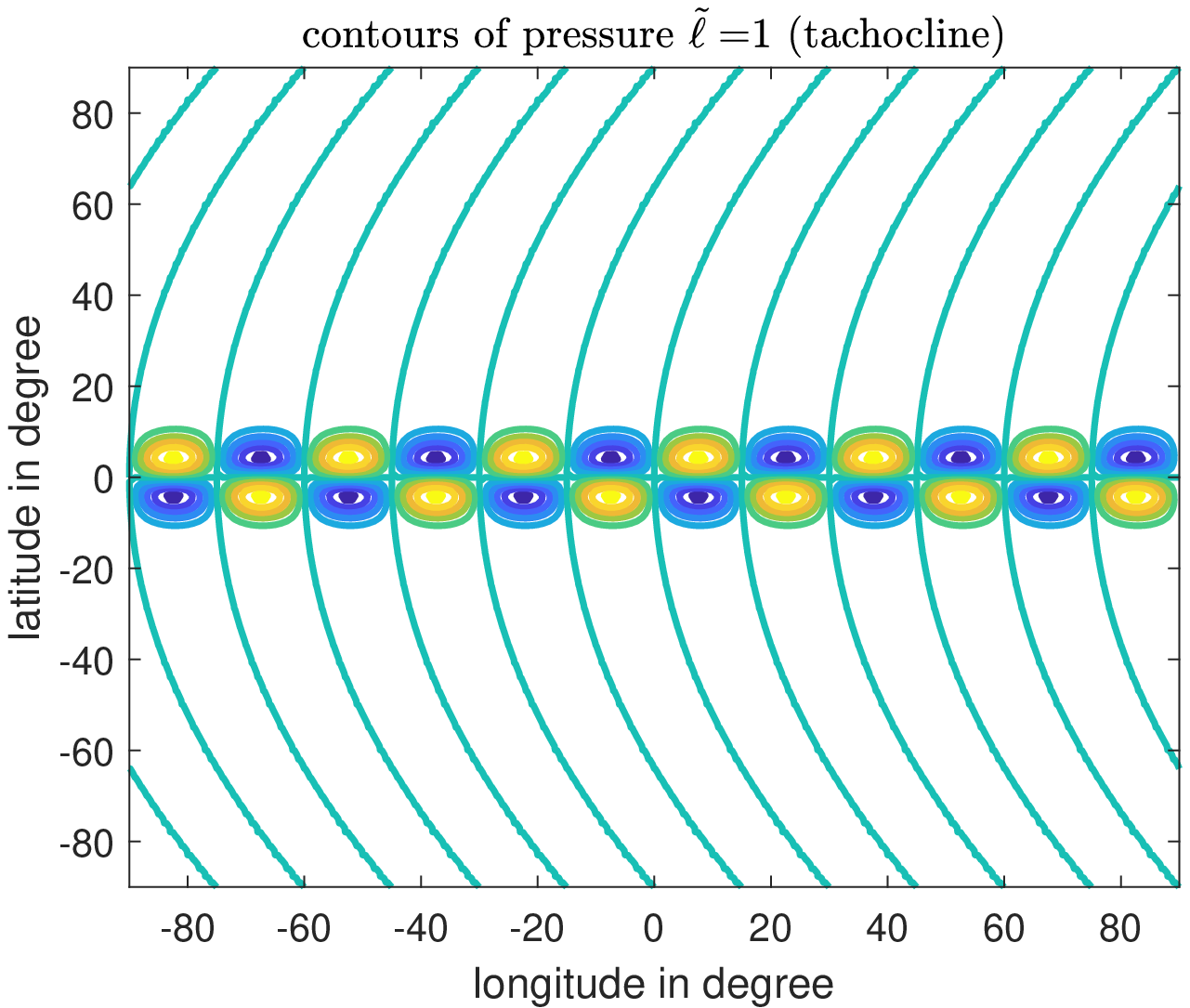}{0.5\textwidth}{(c)}
          \fig{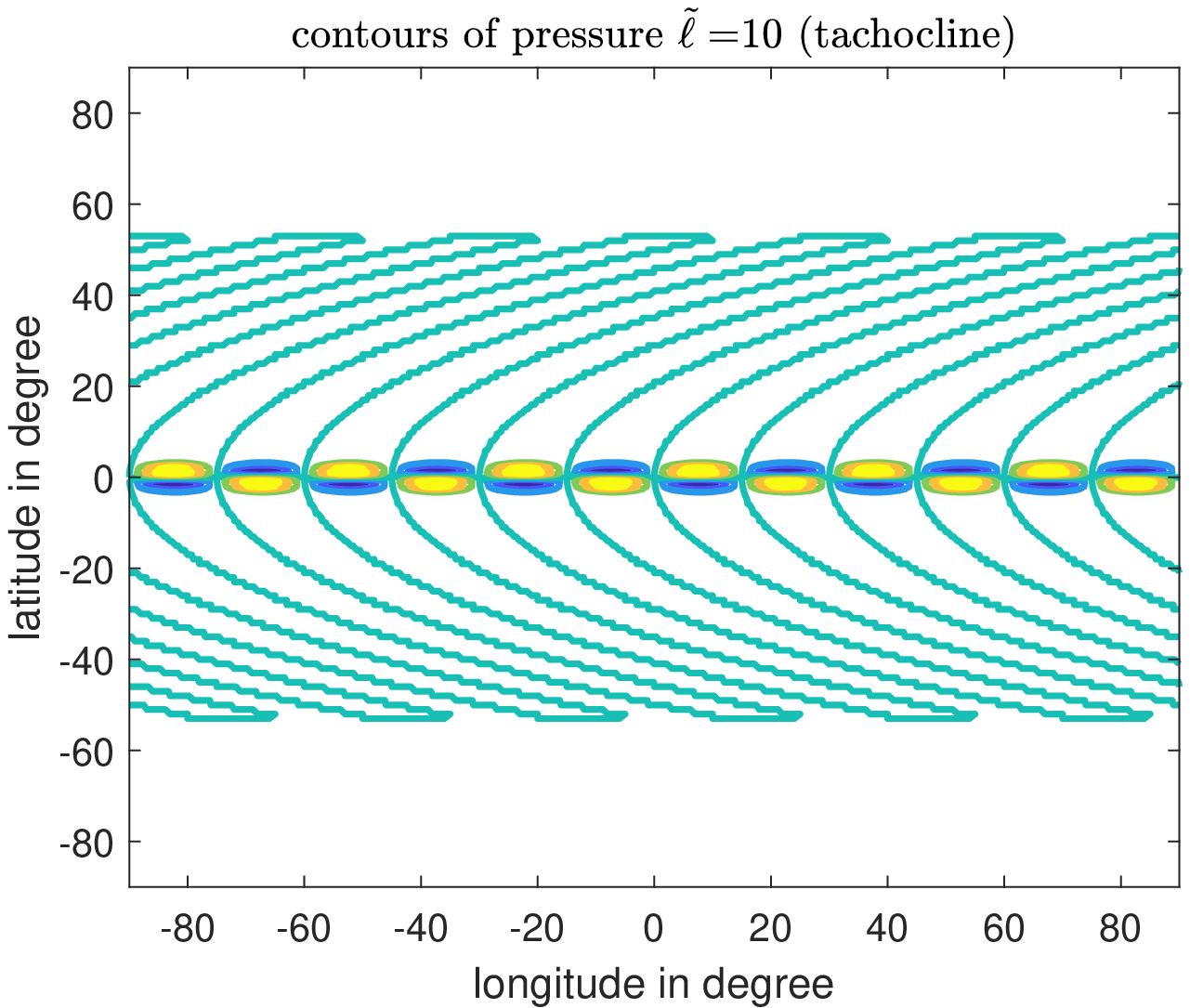}{0.5\textwidth}{(d)}
          }
\caption{The contours of pressure for the mixed Rossby-Poincar\'e waves. (a and b) waves in the solar photosphere for $\tilde{\ell}=1$ and $\tilde{\ell}=10$; (c and d)waves in the solar tachocline for $\tilde{\ell}=1$ and $\tilde{\ell}=10$. The global zonal wavenumber is $\tilde{m}=12$, and the wave period is 160 days. \label{fig:f5}}
\end{figure}

In previous discussions, we have mentioned that the non-traditional Coriolis parameter has an effect on the phase curves. If the effect is strong, the contours of the physical variables will be tilted. The relative strength of stratification to rotation in a fluid can be measured by $N/(2\Omega)$. For the sun, the photosphere is strongly stratified, while the overshooting layer in the tachocline is weakly stratified. Fig.~\ref{fig:f5} shows the contours of pressure on the meridional-zonal plane for the mixed Rossby-Poincar\'e waves with different vertical wavenumbers ($\tilde{\ell}=1$ and $\tilde{\ell}=10$)  in the solar atmosphere and solar tachocline. To signify the effect of rotation on wave patterns, we do not include magnetic filed here. It can be seen from the figure that waves could hardly be equatorially trapped in the solar photosphere when $\tilde{\ell}$ is small (see Fig.\ref{fig:f5}(a)). In (\ref{eq41}), we see that real part of the exponential rate is proportional to the vertical wavenumber. The contours of pressure for a larger wavenumber $\tilde{\ell}=10$ has been plotted in Fig.\ref{fig:f5}(b). It clearly shows the waves are trapped near the equatorial regions. We have not observed significant tilt of pressure contours in the solar atmosphere, since the fluid is strongly stratified over there. In the solar tachocline, however, the situation is different. Figs.~\ref{fig:f5}(c) and \ref{fig:f5}(d) show similar pressure contours for the solar tachocline. First, we notice that waves are more likely to be trapped in the equatorial region in the solar tachocline. Compared with panels (a,b), the band widths in panels (c,d) are significantly shrunk. Second, for the solar tachocline, we also find that the pressure contours tilt due to the non-traditional Coriolis effect. It demonstrates that the non-traditional effect has non-negligible effect on the structure of equatorially trapped waves when the fluid is weakly stratified.

\section{Summary}
In this paper, we have investigated the equatorially trapped waves in the convectively coupled rotating flow, with and without background magnetic field. In our model, we have identified the HD and MHD equatorially trapped Poincar\'e waves, Rossby waves, mixed Rossby-Poincar\'e waves, and Kelvin waves. When the rotational effect is weak, the convectively coupled model has no significant difference from the shallow water model. However, when the rotational effect is strong, they can be substantially different. In such case, it has been found that the wave frequencies of these equatorially trapped waves approach to the Brunt-V\"ais\"al\"a frequency in the convectively coupled model. The non-traditional Coriolis parameter has been included in our model, and it has important effect on the wave frequencies and flow structures. When it is included, the width of the equatorially trapped waves are wider, and phase shifts occur for physical variables. One consequence is that the contours of physical variables will be tilted from the zonal directions. Phase shifts also occur among different variables. Without the non-traditional effect, the pressure, meridional velocity, and vertical velocity are quadrature to the zonal velocity. However, if non-traditional effect is taking into account, these variables are out of quadrature to the zonal velocity.

We have also investigated the equatorially trapped MHD waves. We find that weak magnetic field has negligible effect on equatorially trapped waves. However, the effect of strong magnetic field can be significant. In a strong magnetic field, the speeds of all equatorially trapped waves approach to Alfv\'en speed. With an moderate magnetic field, the MHD wave frequencies are modified by magnetic field. The modification depends on the relative importance of Brunt-V\"ais\"al\"a frequency and Alfv\'en wave frequency.

We have applied our model to the solar atmosphere and solar tachocline. In the solar atmosphere, we find that the Rieger and Rieger type periodicities are consistent with the periods of HD Rossby or mixed Rossby-Poincar\'e waves at high zonal wavenumbers $\tilde{m}$. When magnetic field is included, we find that a trend reversal of wave periods occurs at high $\tilde{m}$. At weak magnetic field, the periods of MHD waves generally increases with $\tilde{m}$. However, the trend reverses at strong magnetic field. For the solar atmosphere, we find that the reversal takes place when the magnetic field reaches a critical value of about $5G$. If the Rieger type periodicities are connected to the equatorially trapped MHD waves at high $\tilde{m}$, then we expect that the toroidal magnetic field at the photosphere should be smaller than $5G$. Otherwise, they must be induced by low $\tilde{m}$ equatorially trapped Rossby waves. In the solar tachocline, we find that the periods of equatorially trapped waves decrease with increasing magnetic strength. To observe the 160 days period, the magnetic field must be smaller than $50kG$.

In this paper, we use a simple Boussinesq model to study the equatorial trapped waves in stars. It can be improved in many aspects. Fist, there are many modes of equatorially trapped waves, but we did not explain why only a few modes were observed. As explained in \citet{zaqarashvili10}, solar differential rotation could possibly play an important role. Second, in rapidly rotating stars or planets, such as observed in Jupiter and Saturn, zonal flows are usually separated into alternative bands. The background zonal shear flows might have important effect on the equatorial waves. It would be interesting to include a background zonal shear in the model. Third, we only consider a constant stratification across the layer. It would also be interesting to discuss the wave behaviors with varying stratifications.

\acknowledgements
We thank the anonymous reviewer for constructive comments, which led to a major improvement of the manuscript. We also thank Zhang Q.-S. for kindly providing us with the data of their latest solar model. T.C. has been supported by NSFC (No.11503097), the Guangdong Basic and Applied Basic Research Foundation (No.2019A1515011625), the Science and Technology Program of Guangzhou (No.201707010006), the Science and Technology Development Fund, Macau SAR (Nos.0045/2018/AFJ, 0156/2019/A3), and the China Space Agency Project (No.D020303). C.Y. has been supported by the National Natural Science Foundation of China (grants 11373064, 11521303, 11733010, 11873103), Yun-nan National Science Foundation (grant 2014HB048), and Yunnan Province (2017HC018). X.W. has been supported by National Natural Science Foundation of China (grant no.11872246) and Beijing Natural Science Foundation (grant no. 1202015).

\end{document}